\documentclass[12pt,a4paper]{article}

\usepackage{amsmath}
\usepackage{amssymb}
\usepackage{epsfig}
\usepackage{graphicx}
\usepackage{cite}
\usepackage{mcite}
\usepackage{textcomp}
\usepackage{a4}
\newcommand{\capdef}{}
\newcommand{\mycaption}[2][\capdef]{\renewcommand{\capdef}{#2}%
        \caption[#1]{{\itshape #2}}}
\makeatletter
\renewcommand{\fnum@table}{\textbf{\tablename~\thetable}}
\renewcommand{\fnum@figure}{\textbf{\figurename~\thefigure}}
\makeatother

\newcounter{myenumi}

\renewcommand{\themyenumi}{\roman{myenumi}}
{\end{list}}

\newlength{\myem}
\newcounter{mysubequation}[equation]

\makeatletter
\renewcommand{\section}{\@startsection{section}{1}{0em}{-\baselineskip}%
{\baselineskip}{\normalfont\large\bfseries}}
\renewcommand{\subsection}%
{\@startsection{subsection}{2}{0em}{-0.7\baselineskip}%
{0.7\baselineskip}{\normalfont\bfseries}}
\makeatother

\newcommand{\equ}[1]{\eq~(\ref{equ:#1})}
\newcommand{\figu}[1]{\fig~\ref{fig:#1}}
\newcommand{\bi}{\begin{itemize}}
\newcommand{\ei}{\end{itemize}}

\newcommand{\be}{\begin{equation}}
\newcommand{\ee}{\end{equation}}
\newcommand{\bea}{\begin{eqnarray}}
\newcommand{\eea}{\end{eqnarray}}

\newcommand{\ldm}{\Delta m_{31}^2}
\newcommand{\sdm}{\Delta m_{21}^2}
\newcommand{\deltacp}{\delta_\mathrm{CP}}
\newcommand{\stheta}{\sin^2 2 \theta_{13}}

\newcommand{\ie}{{\it i.e.}}

\newcommand{\eg}{{\it e.g.}}

\newcommand{\cf}{{\it cf.}}

\newcommand{\eq}{Equation}
\newcommand{\eqs}{Equations}

\newcommand{\fig}{Figure}
\newcommand{\Fig}{Figure}

\newcommand{\Ref}{Ref.}
\newcommand{\Refs}{Refs.}
\newcommand{\Sec}{Section}
\newcommand{\Secs}{Sections}
\newcommand{\App}{Appendix}
\newcommand{\Apps}{Appendices}
\newcommand{\Tab}{Table}

\newcommand{\JHFSK}{\mbox{{\sf JHF-SK}}}
\newcommand{\NUMI}{\mbox{{\sf NuMI}}}
\newcommand{\ReactorI}{\mbox{{\sf Reactor-I}}}
\newcommand{\ReactorII}{\mbox{{\sf Reactor-II}}}

\newcommand{\dm}[1]{{\Delta m^2_{#1}}}

\newcommand{\rue}{{\nu_\mu\rightarrow\nu_e}}

\begin{document}
%%%%%%%%%%%%%%%%%%%%%%%%%%%%%%%%%%%%%%%%%%%%%%%%%%%%%%%%%%%%%%%%%%%%%
%%%%                     Title-page                              %%%%
%%%%%%%%%%%%%%%%%%%%%%%%%%%%%%%%%%%%%%%%%%%%%%%%%%%%%%%%%%%%%%%%%%%%%

\begin{titlepage}

% the footnote symbols are only redefined for the title page !
\renewcommand{\thefootnote}{\alph{footnote}}

\vspace*{-3.cm}
\begin{flushright}
TUM-HEP-504/03\\
MPI-PhT/2003-14\\
%hep-ph/0303232
\end{flushright}

\vspace*{0.5cm}

\renewcommand{\thefootnote}{\fnsymbol{footnote}}
\setcounter{footnote}{-1}

{\begin{center} {\large\bf Reactor Neutrino Experiments Compared to
Superbeams$^*$\footnote{\hspace*{-1.6mm}$^*$Work supported by
``Sonderforschungsbereich 375 f\"ur Astro-Teilchenphysik'' der
Deutschen Forschungsgemeinschaft and the ``Studienstiftung des
deutschen Volkes'' (German National Merit Foundation) [W.W.].}}
\end{center}} \renewcommand{\thefootnote}{\alph{footnote}}

\vspace*{.8cm}
\vspace*{.3cm}
{\begin{center} {\large{\sc
                P.~Huber\footnote[1]{\makebox[1.cm]{Email:}
                phuber@ph.tum.de},~
                M.~Lindner\footnote[2]{\makebox[1.cm]{Email:}
                lindner@ph.tum.de},
                T.~Schwetz\footnote[3]{\makebox[1.cm]{Email:}
                schwetz@ph.tum.de},     
                ~and~
                W.~Winter\footnote[4]{\makebox[1.cm]{Email:}
                wwinter@ph.tum.de}~
                }}
\end{center}}
\vspace*{0cm}
{\it
\begin{center}

\footnotemark[1]${}^,$\footnotemark[2]${}^,$\footnotemark[3]${}^,$\footnotemark[4]%
       Institut f\"ur theoretische Physik, Physik--Department,\\
       Technische Universit\"at M\"unchen,
       James--Franck--Strasse, D--85748 Garching, Germany

\footnotemark[1]%
       Max-Planck-Institut f\"ur Physik, Postfach 401212,
       D--80805 M\"unchen, Germany

\end{center}}

\vspace*{1.5cm}

{\Large \bf
\begin{center} Abstract \end{center}  }

We present a detailed quantitative discussion of the measurement of the leptonic mixing angle $\sin^2 2 \theta_{13}$ with a future reactor neutrino oscillation experiment consisting of a near and far detector. We perform a thorough analysis of the impact of various systematical errors and compare the resulting physics  potential to the one of planned first-generation superbeam experiments. Furthermore, we investigate the complementarity of both types of experiments. We find that, under realistic assumptions, a determination of $\sin^2 2 \theta_{13}$ down to $10^{-2}$ is possible with reactor experiments. They are thus highly competitive to first-generation superbeams and may be able to test $\sin^2 2 \theta_{13}$ on shorter timescales.  In addition, we find that the combination of a KamLAND-size reactor experiment with one or two superbeams could substantially improve the ability to access the neutrino mass hierarchy or the leptonic CP phase.

\vspace*{.5cm}

\end{titlepage}

\newpage

\renewcommand{\thefootnote}{\arabic{footnote}}
\setcounter{footnote}{0}

%%%%%%%%%%%%%%%%%%%%%%%%%%%%%%%%%%%%%%%%%%%%%%%%%%%%%%%%%%%%%%%%%%%%%
%                     Main text                                     %
%%%%%%%%%%%%%%%%%%%%%%%%%%%%%%%%%%%%%%%%%%%%%%%%%%%%%%%%%%%%%%%%%%%%%

\section{Introduction}

There exists now strong evidence for
atmospheric~\cite{Fukuda:1998mi,*Ambrosio:1998wu,*Ronga:2001zw,SK_atm_nu2002}
and solar neutrino
oscillations~\cite{Cleveland:1998nv,*Abdurashitov:2002xa,%
*Hampel:1998xg,*Altmann:2000ft,*Fukuda:2002pe,*Ahmad:2002jz,*Ahmad:2002ka}.
The recent KamLAND reactor
experiment~\cite{Eguchi:2002dm} together with all the solar neutrino
data clearly confirms the LMA solution for the mass
splittings and mixings and excludes other leading flavor
transition mechanisms (for a recent review
see, for example, \Ref~\cite{Pakvasa:zv}).  Furthermore, the CHOOZ
reactor experiment~\cite{Apollonio:1999ae,Apollonio:2002gd} currently
provides the most stringent upper bound for the small sub-leading parameter
$\stheta$. Since a non-vanishing value of $\stheta$ is a prerequisite
for genuine three-flavor effects, such as leptonic CP
violation, it is now one of the most important challenges to establish
$\stheta > 0$ or at least to improve the sensitivity limit. Some
improvement on $\stheta$ may be obtained from conventional beam
experiments, such as the ongoing K2K
experiment~\cite{Ahn:up}, or the MINOS~\cite{Paolone:2001am} and CNGS~\cite{Duchesneau:2002yq} experiments under construction~\cite{Migliozzi:2003pw}. Even
better limits can be obtained with planned
superbeams~\cite{Itow:2001ee,Ayres:2002nm,Asratyan:2003dp,Huber:2002rs,Whisnant:2002fx,Minakata:2001qm,Barger:2000nf,*Gomez-Cadenas:2001eu,*Aoki:2001rc,*Aoki:2002ks,*Barenboim:2002zx,*Aoki:2002ae,*Barenboim:2002nv,*Okamura:2002pb,*Mezzetto:2003mm,*Diwan:2003bp} or
future neutrino factory experiments~(for a summary, see
\Ref~\cite{Apollonio:2002en}). Especially superbeam experiments,
such as the JHF to Super-Kamiokande~\cite{Itow:2001ee} and the
NuMI~\cite{Ayres:2002nm} experiments, are being planned with the main
purpose to measure $\stheta$. Both superbeam and neutrino
factory experiments suffer, however, from the presence of parameter correlations and degeneracies~\cite{Fogli:1996pv,Burguet-Castell:2001ez,Minakata:2001qm,Barger:2001yr}, and the $\stheta$ measurements would not be as good as expected from
statistics and systematics only~\cite{Huber:2002mx}. Therefore,
various suggestions have been made to resolve the correlations and
degeneracies by the combination of at least two long-baseline
experiments~\cite{Asratyan:2003dp,Huber:2002rs,Whisnant:2002fx,Barger:2002rr,*Huber:2003ak,*Burguet-Castell:2002qx,*Minakata:2002qi,Barger:2002xk,*Minakata:2003ca}.

Oscillation experiments at nuclear power plants have recently been
suggested as an interesting alternative to superbeams for the
measurement of
$\stheta$~\cite{Mikaelyan:1999pm,*Mikaelyan:2000st,*Mikaelyan:2002nv,Martemyanov:2002td,Minakata:2002jv,reactor_US}. Reactor
experiments have a long history in neutrino physics. Starting from the
legendary Cowan-Reines experiment~\cite{Cowan:1956xc}, many
measurements at nuclear power plants have provided valuable
information about neutrinos. Very important are the results of the
G{\"o}sgen~\cite{Zacek:1986cu}, Bugey~\cite{Declais:1995su}, Palo
Verde~\cite{Boehm:2001ik}, and CHOOZ~\cite{Apollonio:1999ae}
experiments, which have lead to stringent limits on electron
antineutrino disappearance. Reactor neutrino experiments have become
very prominent again due to the outstanding results of the KamLAND
experiment~\cite{Eguchi:2002dm}. For a recent review on reactor
neutrino experiments, see \Ref~\cite{Bemporad:2001qy}.

In order to improve the information on $\stheta$, it has been proposed to use a
reactor experiment with a near detector very close to the reactor
complex and a far detector at a distance of $\sim 1 \, \mathrm{km} -2
\, \mathrm{km}$. Systematical errors can be reduced in this way, and a
sensitivity to $\stheta$ down to $\simeq 0.02$ might be
reachable~\cite{Martemyanov:2002td,Minakata:2002jv,reactor_US}.
Moreover, such a measurement would not be spoilt
by correlations and degeneracies. In this study, we analyze such reactor experiments and compare their physics potential to the one of first-generation superbeams. For that purpose, we thoroughly investigate the impact of various systematical errors. In addition, performing a separate and combined
analysis of reactor and superbeam experiments in the general three
flavor framework, we study the competitiveness and the complementarity
of the physics potentials of these types of experiments.

The outline of the paper is as follows. In \Sec~\ref{sec:framework},
we present the framework of neutrino oscillations and give a
qualitative discussion of the $\stheta$ measurement at reactor and
superbeam experiments using analytical formulas for the oscillation
probabilities. In \Sec~\ref{sec:experiments}, we describe in detail
how we simulate the experiments. In \Sec~\ref{sec:reactor_only}, we
present our results on the sensitivity limit on $\stheta$ obtainable
at a reactor including a thorough discussion of various systematical
uncertainties. In \Sec~\ref{sec:RvsSB}, we compare the limit as well
as the accuracy for $\stheta$ obtainable at reactors with the one of
superbeam experiments, whereas in \Sec~\ref{sec:complementarity} we
show how a combined analysis of reactor and superbeam experiments can
improve significantly the possibilities to identify the neutrino mass
hierarchy and to discover leptonic CP violation. Our conclusions are
presented in \Sec~\ref{sec:conclusion}. Details of the statistical
analysis are given in \App~\ref{app:syst+ND}, and in
\App~\ref{app:reactor} we discuss experimental details of reactor
experiments, including a summary of the key assumptions adopted in
this work. In \App~\ref{app:ND} we investigate the impact of the
position of the near detector on the obtainable sensitivity.  For
readers mainly interested in our physics results, we recommend to
proceed directly to \Secs~\ref{sec:limit}, \ref{sec:RvsSB},
\ref{sec:complementarity}, and~\ref{sec:conclusion}.

\section{The framework of neutrino oscillations}
\label{sec:framework}

Our results are based on a complete numerical three-flavor analysis. However, it is useful to have a qualitative analytical understanding
of most effects. Therefore, we expand the relevant oscillation
probabilities in terms of the small mass hierarchy parameter $\alpha
\equiv \sdm/\ldm$ and the small mixing angle $\sin 2\theta_{13}$ using
the standard parameterization of the leptonic mixing matrix
$U$~\cite{PDG}. For the superbeams considered in this paper, we use,
for the sake of simplicity, the expansion from
\Refs~\cite{Freund:2001ui,CERVERA,FREUND} in vacuum, which is at least
a good approximation for the JHF to Super-Kamiokande
experiment.\footnote{For similar expansions in matter, see, for
example, \Refs~\cite{FLPR,CERVERA,FREUND}.} For the
appearance signal with terms up to the second order, \ie, proportional
to $\sin^2 2\theta_{13}$, $\sin 2\theta_{13} \cdot \alpha$, and
$\alpha^2$, one has
\begin{eqnarray}
P_{\mu e} & \simeq & \sin^2 2\theta_{13} \, \sin^2 \theta_{23}
\sin^2 {\Delta_{31}} \nonumber \\
&\mp&  \alpha\; \sin 2\theta_{13} \, \sin\deltacp  \, \cos\theta_{13} \sin
2\theta_{12} \sin 2\theta_{23}
\sin^3{\Delta_{31}} \nonumber \\
&-&  \alpha\; \sin 2\theta_{13}  \, \cos\deltacp \, \cos\theta_{13} \sin
2\theta_{12} \sin 2\theta_{23}
 \cos {\Delta_{31}} \sin^2 {\Delta_{31}} \nonumber  \\
&+& \alpha^2 \, \cos^2 \theta_{23} \sin^2 2\theta_{12} \sin^2 {\Delta_{31}},
\label{equ:PROBVACUUM}
\end{eqnarray}
where $\Delta_{ij} \equiv \Delta m_{ij}^2 L/(4 E) \equiv (m_i^2-m_j^2)
L/(4E)$, and the sign of the second term refers to neutrinos (minus) or antineutrinos (plus). Depending on the actual values of $\alpha$ and $\sin 2
\theta_{13}$, each of the individual terms in \equ{PROBVACUUM} gets
a relative weight. Thus, since within the LMA-allowed allowed region
this relative weight can favor different terms in \equ{PROBVACUUM}, it
turns out that the $\stheta$-$\sdm$-plane is appropriate to illustrate
the experimental potential to measure $\stheta$, the mass hierarchy,
and CP effects. For example, for large $\alpha$ and large $\stheta$
the second and third terms are favored, which means that CP
measurements become possible especially for large $\sdm$. On the
other hand, for small $\alpha$ the first term in \equ{PROBVACUUM} can
be measured in a clean way without being affected by the other
terms. Therefore, $\stheta$ or the sign of $\ldm$, entering the first
term through matter effects, can be accessed there. In fact, it can be
shown that a simultaneous measurement of $\deltacp$ and the sign of
$\ldm$ independent of the true value of $\sdm$ is hardly possible for
the first-generation superbeams or their
combination~\cite{Huber:2002rs}. The main reason for that problem
is that superbeams suffer from parameter correlations and
degeneracies coming from the different combinations of parameters in
\equ{PROBVACUUM}. Degeneracies are defined as solutions in parameter
space disconnected from the best-fit region at the chosen confidence
level. Many of the degeneracy problems originate in the summation of
the four terms in \equ{PROBVACUUM} especially for large $\alpha$, since changes of one parameter value can be often compensated by adjusting
another one in a different term.  This leads to the
$(\delta, \theta_{13})$~\cite{Burguet-Castell:2001ez},
$\mathrm{sgn}(\Delta m_{31}^2)$~\cite{Minakata:2001qm}, and
$(\theta_{23},\pi/2-\theta_{23})$~\cite{Fogli:1996pv} degeneracies,
\ie, and overall ``eight-fold'' degeneracy~\cite{Barger:2001yr}. For superbeams, the $(\delta, \theta_{13})$-degeneracy does no appear
as a disconnected solution. In addition, we choose the atmospheric best-fit
value $\theta_{23}=\pi/4$, which means that especially the $\mathrm{sgn}(\Delta
m_{31}^2)$-degeneracy will affect our results. We include the
correlations and degeneracies, unless otherwise stated, in our
results, and discuss their influence where appropriate. A detailed
illustration of different sources of measurement errors and their
impact can, for example, be found in \Ref~\cite{Huber:2002mx}. In
addition, the role of the degeneracies and the potential to resolve
them has, for example, been studied in
\Refs~\cite{Asratyan:2003dp,Huber:2002rs,Whisnant:2002fx,Barger:2002xk,Barger:2002rr}.

For the reactor experiments, we can, for short baselines, safely
neglect matter effects. We find, up to second order
in $\sin 2\theta_{13}$ and $\alpha$,
\be
1 - P_{\bar{e} \bar{e}} \quad \simeq \quad \sin^2 2 \theta_{13} \, \sin^2 \Delta_{31} +  \alpha^2 \, \Delta_{31}^2 \, \cos^4 \theta_{13} \, \sin^2 2 \theta_{12} \,.
\label{equ:PROBREACTOR}
\ee
At the first atmospheric oscillation maximum, $\Delta_{31}$ is
approximately $\pi/2$ and $\sin^2 \Delta_{31}$ is close to one, which
means that the second term on the right-hand side of this equation is also very small for $\stheta \gtrsim 10^{-3}$ and can for many purposes be neglected. The reactor measurement is at short baselines for large enough
$\stheta$ therefore dominated by the product of $\stheta$ and $\sin^2
\Delta_{31}$, which must be measured as deviation from one. The simple structure of \equ{PROBREACTOR} implies that, in comparison to the superbeams, correlations and degeneracies only play a minor role in reactor experiments -- they are almost
exclusively dominated by statistical and systematical errors. In other
words, they can be used as ``clean laboratories for $\theta_{13}$
measurements''~\cite{Minakata:2002jv}. Especially, the behavior in
the $\stheta$-$\sdm$-plane will be different to the one for
superbeams, since \equ{PROBREACTOR} is almost independent of $\sdm$. The
direct comparison of reactor experiments and superbeams with respect
to the most important dependencies will be one of the main aspects of
this study. \equ{PROBREACTOR} also demonstrates the limitations of reactor experiments, since there is no dependence on
$\theta_{23}$, $\deltacp$, and the sign of $\ldm$ in this
formula. However, as we will show, the sensitivity to the sign of
$\ldm$ and to CP violation of superbeam experiments can be
significantly improved by the {\it combined} analysis of reactor and
superbeam experiments, through the precise determination of $\stheta$
at the reactor.
 
All results within this study are, unless otherwise stated, calculated
for the following values for the oscillation parameters, given with
the currently allowed $3 \sigma$-ranges:
\begin{eqnarray}
\ldm & =& 3_{-2}^{+3}\cdot10^{-3}\,\mathrm{eV}^2\,, \nonumber \\
\sin^2 2\theta_{23} & = & 1_{-0.2}^{+0}\,, \nonumber \\
\sdm & = & 7_{-3}^{+23}\cdot10^{-5}\,\mathrm{eV}^2\,, \nonumber \\
\sin^2 2\theta_{12} & = & 0.8_{-0.2}^{+0.2}\,. 
\label{equ:params}
\end{eqnarray}
These numbers are motivated by recent global fits to atmospheric plus
K2K data, such as in
\Refs~\cite{SK_atm_nu2002,Ahn:up,Gonzalez-Garcia:2002mu,*Maltoni:ni,*Fogli:th},
and KamLAND plus solar neutrino experiments, such as in
\Ref~\cite{Maltoni:2002aw,*Bahcall:2002ij,*Fogli:au,*deHolanda:2002iv,*Bandyopadhyay:2002en}. Throughout
this work, we refer to the solar parameters given in \equ{params} as the
LMA solution. In some cases, we will differentiate between the
so-called LMA-I and LMA-II solutions with the best-fit values $\sdm =
7 \cdot 10^{-5} \, \mathrm{eV}^2$ and $\sdm = 1.4 \cdot 10^{-4} \,
\mathrm{eV}^2$, respectively, which both are covered by the
$3\sigma$-allowed range in \equ{params}. Unless otherwise stated, we
assume a normal mass hierarchy, \ie, $\Delta m_{31}^2 = + 3.0 \cdot
10^{-3} \, \mathrm{eV}^2$. For $\sin^2 2\theta_{13}$, we only allow
values below the CHOOZ bound~\cite{Apollonio:1999ae}, \ie, $\sin^2 2
\theta_{13} \lesssim 0.1$.  For the CP phase, we do not make specific
assumptions, \ie, we allow any value between $0$ and $2 \pi$.

\section{The experiments and their analysis}
\label{sec:experiments}

In order to reliably assess the physics potential of a given
experiment, a realistic event rate calculation in connection with a
proper statistical treatment of the simulated data is needed.\footnote{Further details about the general calculation and analysis techniques used in this work can be found in \Ref~\cite{Huber:2002mx}.} The calculation of event rates is basically a convolution of the flux spectrum with the cross sections, the oscillation probability, the detector efficiency, and the detector energy response function. The resulting events rates are then the basis for a $\chi^2$-analysis,
where systematical uncertainties, correlations and degeneracies are
properly included. The analysis for the superbeam experiments is done as in \Refs~\cite{Huber:2002mx,Huber:2002rs}. It is based on a
Poissonian $\chi^2$, since the appearance channel may have very low
event numbers.  On the other hand, the event rates in the disappearance channel of reactor experiments are quite large and we can use a Gau\ss ian
$\chi^2$, which has the advantage of
allowing a more transparent inclusion of systematical errors. For the
detailed definition of the $\chi^2$ functions, we refer to
\App~\ref{app:syst+ND}. Eventually, the combined analysis of reactor
and superbeam experiments is done in a similar way as in
\Ref~\cite{Huber:2002rs}, where we pay special attention to the
correct treatment of degeneracies. All of the figures in this work
are shown at the $90\%$ confidence level.

Once we have obtained a $\chi^2$-value including the effects of
systematical uncertainties, we project onto the parameter of interest
in the six-dimensional space of the oscillation parameters $\ldm$,
$\sdm$, $\theta_{12}$, $\theta_{13}$, $\theta_{23}$, and
$\deltacp$. In addition, we take into account external information
coming from other experiments, \ie, we assume to know the parameters
$\ldm$, $\sin^2 2 \theta_{12}$, and $\sdm$ with a $10\%$
$1\sigma$-error from K2K~\cite{Ahn:up}, MINOS~\cite{Paolone:2001am},
CNGS~\cite{Duchesneau:2002yq}, and
KamLAND~\cite{Eguchi:2002dm,BARGER,*deGouvea:2001su,*Gonzalez-Garcia:2001zy}
results. This should be realistic at the time when the reactor
or superbeam analysis will be performed. It turns out that
our results are however rather insensitive towards the precise values of those errors as long as they are reasonably small, \ie, below about $30 \%$ to $50\%$.
The matter density for the superbeam experiments is assumed to be
known to $5\%$~\cite{Geller:2001ix}, which is essentially only
relevant for the \NUMI\ experiment.  In the following two
subsections, we provide further details on the simulations of the reactor
experiments with near and far detectors and superbeams.

\subsection{Future reactor experiments with near and far detectors}
\label{sec:exp_reactors}

Nuclear fission reactors are a strong and pure source of low energy 
$\bar\nu_e$. Appearance experiments are not possible due to the low energies, and  the inverse $\beta$--decay with an energy threshold of $1.804\,\mathrm{MeV}$ is by far the dominant detection reaction:
\begin{equation}
\bar\nu_e + p \rightarrow e^+ + n\,.
\end{equation}
This reaction has a very distinctive experimental signature which consists of
the $\gamma$-rays from the annihilation of the positron and of the delayed 
signal of the neutron capture. This delayed coincidence allows to reject most of the background events. The relation between the positron energy and the
neutrino energy is given by
\begin{equation}
\label{eq:energy}
E_{\bar\nu_e}=E_{e^+} +(m_n-m_p)+O(E_{\bar\nu_e}/m_n)\, ,
\end{equation}
where $E_{e^+}$ is the sum of the kinetic energy of the positron and
the rest mass of the positron. The energy visible in the detector is
determined by $E_\mathrm{vis}=E_{e^+}+511\,\mathrm{keV}$, where the
additional $511\,\mathrm{keV}$ come from the mass of the electron
which with the positron annihilates.  Thus, a neutrino with the
threshold energy produces already $2\cdot511\, \mathrm{keV}$ of
visible energy.  A precise measurement of the visible energy
$E_{\mathrm{vis}}$ yields according to \eq~\ref{eq:energy} a unique
determination of the neutrino energy $E_{\bar\nu_e}$ (for details of
the energy reconstruction, see \App~\ref{app:reactor}). The cross
section for inverse $\beta$-decay has approximately the form
\begin{equation}
\sigma(E_{e^+})\simeq\frac{2\pi^2\hbar^3}{m_e^5 f \tau_n}p_{e^+}E_{e^+}\,,
\end{equation}
where $\tau_n$ is the lifetime of a free neutron and $f$ is the free
neutron decay phase space factor. In our numerical calculations we
use the cross sections from \Ref~\cite{Vogel:1999zy} including higher
order corrections. For the neutrino flux spectrum from a nuclear
power reactor we use the parameterization of
\Ref~\cite{Vogel:1989iv,*Murayama:2000iq}, and we adopt a fuel
composition as in \Ref~\cite{Eguchi:2002dm}.

\begin{figure}[ht!]
\begin{center}
\includegraphics[width=16cm]{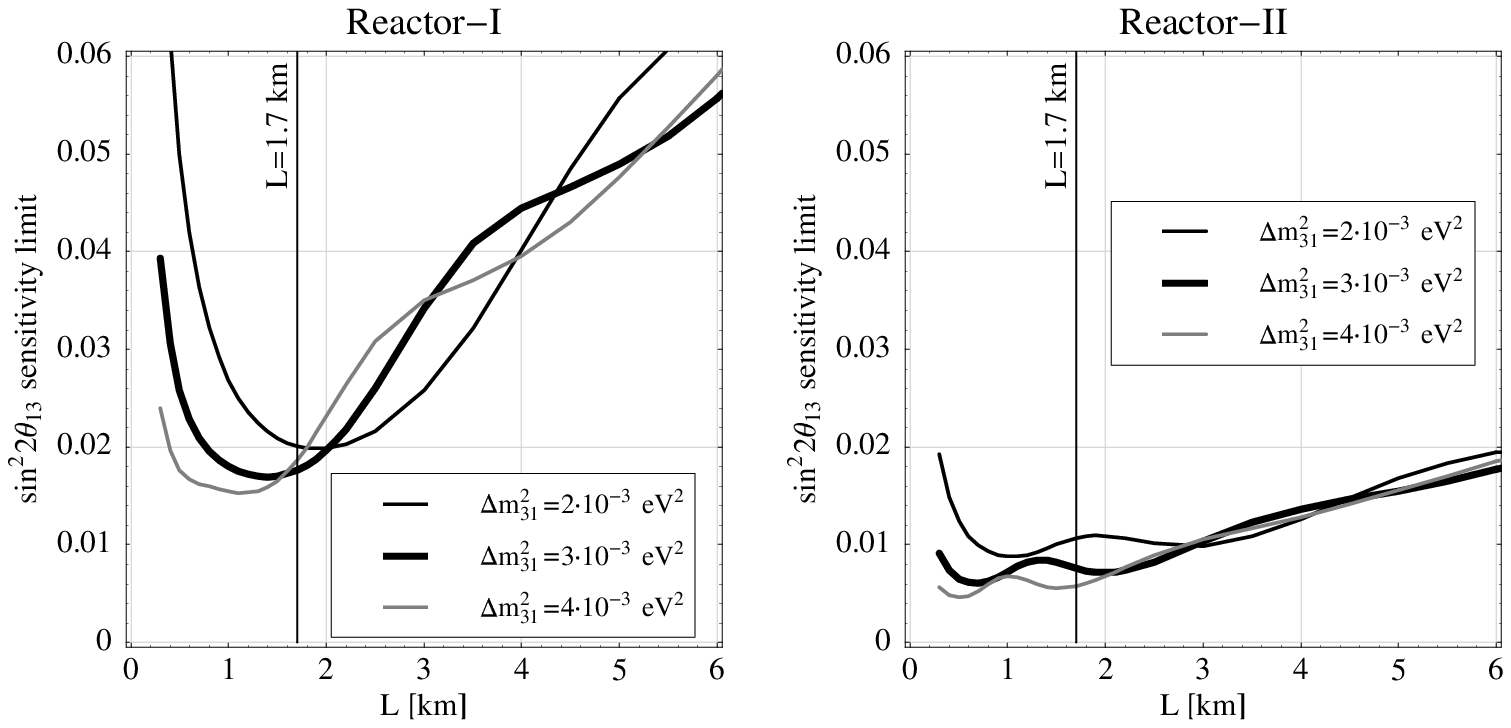}
\end{center}
\mycaption{\label{fig:basedep} The $\sin^2 2 \theta_{13}$ sensitivity limit as function of the baseline $L$ for the \ReactorI\ (left plot) and \ReactorII\ (right plot) setups. It is shown for several values for $\ldm$ at the 90\% confidence level. The vertical lines mark the baseline we are using for our standard setups.}
\end{figure}

We assume a detector technology very similar to the
CHOOZ~\cite{Apollonio:2002gd} and KamLAND~\cite{Eguchi:2002dm}
detectors, as well as the Borexino counting test
facility~\cite{Alimonti:1998nt} (\cf, \App~\ref{app:reactor}). In this
work, we suggest a reactor experiment designed to measure
$\theta_{13}$. This measurement is based on the ability to detect
small spectral distortions in the positron event rates due to neutrino
oscillations. For a sensitivity to $\sin^22\theta_{13}$ of the order
of $10^{-2}$ this requires a measurement accuracy of about $1\%$. The
total experimental uncertainties are, however, considerably larger and
come from various sources, such as reactor burn up effects, average
neutrino yields, and cross section. In order to control these
uncertainties, it is necessary to use a near detector, which
accurately determines the energy dependence and total normalization of
the neutrino flux~\cite{Martemyanov:2002td,Minakata:2002jv}. The
details of the near detector simulation are described in
\Apps~\ref{app:syst+ND} and \ref{app:ND}. We furthermore do not
include backgrounds since, as we discuss in detail in
\App~\ref{app:reactor}, they can be suppressed to a negligible
level~\cite{Schonert:2002ep}.

We do not propose a specific reactor site in this study and for
simplicity we assume a single reactor block. For a given site with
several reactors effects like different distances of the cores to near
and far detectors or on-off times of the individual reactors have to
be included. However, one expects that the main results of this work
will also apply in such more complicated situations. We define the
integrated luminosity ${\cal L}$ of a reactor experiment in units of
fiducial detector mass [tons] $\times$ thermal reactor power [GW] $\times$
running time [years].\footnote{Note that our definition of the integrated luminosity assumes a $100\%$ detection efficiency and no deadtimes. In addition, it includes the fiducial detector mass, not the total detector mass. Thus,
for a specific setup, the losses due to these factors imply a rescaling of the luminosity.} In \fig~\ref{fig:basedep}, the dependence of the
$\stheta$ sensitivity on the baseline of the far detector is shown for
two values of the integrated luminosity $400\,
\mathrm{t}\,\mathrm{GW}\,\mathrm{y}$ (left-hand panel) and $8000\,
\mathrm{t}\,\mathrm{GW}\,\mathrm{y}$ (right-hand panel), as well as
several values of $\ldm$. It turns out that a baseline of
$1.7\,\mathrm{km}$ performs reasonably well for all of the considered
cases, which is in agreement with \Ref~\cite{Minakata:2002jv}. This
choice is also best for the given uncertainty of $\ldm$ within the
atmospheric allowed region, as it lies very close to the optimum for
all values of $\ldm$. We assume that the near detector is located at
around $0.17\,\mathrm{km}$ in order to avoid oscillation effects. At
those very short distances, the contribution of other power stations
is far below $1\%$ of the total rate. As long their contribution is
known better than to about $10\%$, they do therefore not contribute to
the total error to more than $0.1\% $ and we can safely ignore them.

\begin{table}[h!]
\begin{center}
\begin{tabular}{|l|c|c|}
\hline
&\ReactorI\ &\ReactorII\ \\
\hline
Integrated luminosity& $400\,
\mathrm{t}\,\mathrm{GW}\,\mathrm{y}$& $8000\,
\mathrm{t}\,\mathrm{GW}\,\mathrm{y}$\\
Unoscillated events&$31\,493$&$629\,867$\\
$\sigma_\mathrm{norm}$&$0.8\%$&$0.8\%$\\
$\sigma_\mathrm{cal}$&$0.5\%$&$0.5\%$\\
Baseline&$1.7\,\mathrm{km}$&$1.7\,\mathrm{km}$\\
\hline
\end{tabular}
\end{center}
\mycaption{\label{tab:reactor} The most important parameter values of the two reactor benchmark setups \mbox{\ReactorI} and \ReactorII\ used in this work. For more information see \App~\ref{app:reactor}.}
\end{table}

Throughout this work, we use two experimental benchmark setups labeled
as \ReactorI\ and \ReactorII , which are defined in
\Tab~\ref{tab:reactor}. The two values of the integrated luminosity in
the table reflect the range of possible experiments with detector
masses of the order of
$10\,\mathrm{t}$~\cite{Martemyanov:2002td,Minakata:2002jv} up to the
order of
$>100\,\mathrm{t}$~\cite{Schonert:2002ep,Eguchi:2002dm}\footnote{Typical
reactors have a thermal power $\sim 2 \, \mathrm{GW}$. Note, however,
that reactor stations with a total thermal power of up to $\sim 24 \,
\mathrm{GW}$ exist~\cite{Minakata:2002jv}.}. The near and far
detectors are assumed to be identical (maybe apart from their size) in
order to minimize the impact of systematical errors. We assume as our
standard value for the uncertainty on the event normalization
$\sigma_\mathrm{norm}=0.8\%$~\cite{Minakata:2002jv}. This has to be
considered as an effective error, receiving contributions from
individual uncertainties of the near and far detectors, as well as
from over-all flux uncertainties (for details, see
\Sec~\ref{sec:systematics} and \App~\ref{app:syst+ND}).  In addition,
we include the effect of an energy calibration error
$\sigma_\mathrm{cal}=0.5\%$~\cite{Apollonio:2002gd,oberauer}

\subsection{First-generation Superbeams}
\label{sec:exp_superbeams}

The two superbeam experiments considered in this study are the JHF to Super-Kamiokande \cite{Itow:2001ee} and the
{\sf NuMI}~\cite{Ayres:2002nm} proposals, referred to as
``{\sf JHF-SK}'' and ``{\sf NuMI}''. Both 
projects will use the $\rue$ appearance channel. 
Neutrino beams which are produced by meson
decays always contain irreducible fractions of $\nu_e$, $\bar\nu_e$, and $\bar\nu_\mu$ contaminations, as well as they have a large high energy tail. Both experiments use therefore an off-axis setup to make the spectrum
much narrower in energy and to suppress the beam contaminations~\cite{offaxis}.
An off-axis beam reaches in this way the low background levels necessary for a good sensitivity to the $\rue$ appearance signal. Both experiments are planned to be operated at nearly the same $L/E$,
which is optimized for the first maximum of the atmospheric oscillation for $\dm{31}=3.0\cdot10^{-3}\,\mathrm{eV}^2$.

\begin{table}[ht!]
\begin{center}
\begin{tabular}{|l|c|c|}
\hline
&{\sf JHF-SK}&{\sf NuMI}\\
\hline
\multicolumn{3}{|l|}{Beam} \\
\hline
Baseline&$295\,\mathrm{km}$&$712\,\mathrm{km}$\\
Target Power& $0.77 \, \mathrm{MW}$ & $0.4 \, \mathrm{MW}$ \\
Off-axis angle&$2^\circ$&$0.72^\circ$\\
Mean energy&$0.76\,\mathrm{GeV}$&$2.22\,\mathrm{GeV}$\\
Mean
$L/E$&$385\,\mathrm{km}\,\mathrm{GeV}^{-1}$&$320\,\mathrm{km}\,\mathrm{GeV}^{-1}$\\
\hline
\multicolumn{3}{|l|}{Detector} \\
\hline
Technology&Water Cherenkov&Low-Z calorimeter\\
Fiducial mass&$22.5\,\mathrm{kt}$&$17\,\mathrm{kt}$\\
Running period&5 years&5 years\\
\hline
\end{tabular}
\end{center}
\mycaption{\label{tab:base} The two superbeams and their detectors as given
in \Refs~{\rm \protect\cite{Itow:2001ee,Ayres:2002nm}} }
\end{table}

Due to the different energies of the two beams, different detector
technologies are used. For the JHF beam, Super-Kamiokande, a water
Cherenkov detector with a fiducial mass of $22.5\,\mathrm{kt}$, is assumed.
The Super-Kamiokande detector has an excellent electron muon separation
and NC (neutral current) rejection. For the {\sf NuMI} beam, a low-Z calorimeter with a fiducial
mass of $17\,\mathrm{kt}$ is planned, because the hadronic fraction of the
energy deposition is much larger at those energies. In spite of the very
different detector technologies, their performances in terms of background levels and
efficiencies are rather similar. The actual numbers for these quantities and the respective energy
resolution of the detectors can be found in~\cite{Huber:2002rs}. 

In \Ref~\cite{Huber:2002rs}, several modifications of those setups have been 
considered in order to improve either the sensitivity to CP violation or to
the mass hierarchy. There, it is demonstrated that \JHFSK\ with about $1/4$ of the total running time with neutrinos and about $3/4$ with antineutrinos performs best for the CP measurement\footnote{This setup has about equal total numbers of neutrino and antineutrino events.}, a setup which is labeled as \JHFSK $_{cc}$ in this work. However, for the determination of the mass hierarchy the \NUMI\ setup with an increased baseline of $890\,\mathrm{km}$ at an off-axis angle of $0.72^\circ$ performs better because of larger matter effects, which will be labeled as \NUMI @$890\,\mathrm{km}$ in this work.

\section{Measuring $\stheta$ at a reactor, and the impact of systematical
errors}
\label{sec:reactor_only}

In this section, we investigate in detail the potential of a reactor
experiment to measure $\stheta$. In \Sec~\ref{sec:limit}, we discuss
obtainable limits on $\stheta$, and we present in
\Secs~\ref{sec:systematics} and~\ref{sec:cut} a detailed discussion of
systematical errors. The results of these sections are obtained by
neglecting correlations of the oscillation parameters, \ie, all
parameters except from $\stheta$ are fixed to the values given in
\equ{params}. However, as it can be inferred from the discussion
related to \equ{PROBREACTOR} and as we will see explicitly in the
numerical calculations presented in \Secs~\ref{sec:RvsSB}
and~\ref{sec:complementarity}, the influence of other oscillation
parameters on the $\stheta$ sensitivity limit from reactors is very
small.

\subsection{The sensitivity limit to $\stheta$}
\label{sec:limit}

\begin{figure}[t!]
\begin{center}
\includegraphics[angle=-90, width=14cm]{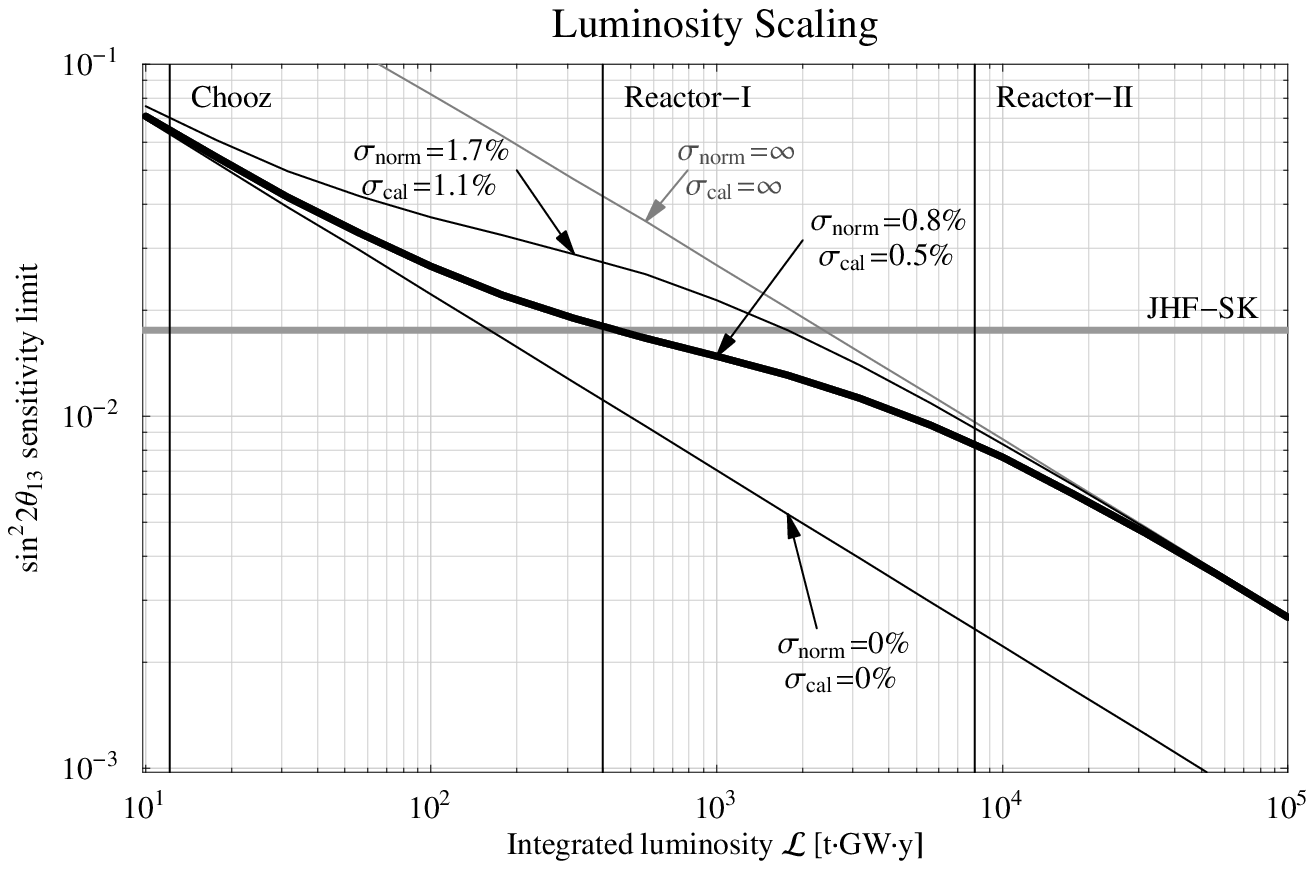}
\end{center}
\mycaption{\label{fig:luminosity} The sensitivity to $\stheta$ as a function of the integrated luminosity for different values of the normalization error $\sigma_\mathrm{norm}$ and the energy calibration error $\sigma_\mathrm{cal}$ at the 90\% confidence level, where the thick curve corresponds to our standard values. The vertical lines mark the luminosities of the CHOOZ experiment and our standard setups \ReactorI\ and \ReactorII\ as defined in \Tab~\ref{tab:reactor}. The horizontal line shows a typical $\stheta$ sensitivity limit obtainable by the  \JHFSK\ superbeam experiment for the same parameter values (taken from \Sec~\ref{sec:RvsSB}).}
\end{figure}

We define the sensitivity or sensitivity limit to $\stheta$ as the
largest value of $\stheta$, which is allowed at a given confidence
level if the true value is $\stheta \equiv 0$.  In
\Fig~\ref{fig:luminosity}, we show the sensitivity to $\stheta$ as a
function of the integrated luminosity ${\cal L}$ in units of detector
mass [tons] $\times$ thermal reactor power [GW] $\times$ running time
[years].  In this figure, the lower diagonal curve corresponds to the
idealized case of statistical errors only, and shows just the expected
$1/\sqrt{\cal L}$ scaling. Our standard values for the errors on the
normalization $\sigma_\mathrm{norm} = 0.8\%$ and energy calibration
$\sigma_\mathrm{cal} = 0.5\%$ lead to the thick curve, which
represents one of the main results of this paper. At a luminosity
around $100 \, \mathrm{t \, GW \, y}$, we detect a departure from the statistics
dominated regime into a flatter systematics dominated region. As discussed in \Secs~\ref{sec:systematics} and \ref{sec:cut},
this effect is dominated by the error on the normalization
$\sigma_\mathrm{norm}$, whereas the energy calibration error
$\sigma_\mathrm{cal}$ only plays a minor role. However, at large
luminosities $\gtrsim 10^4 \, \mathrm{ t \, GW \, y}$, the slope of the curve changes, and we are entering again a statistics dominated region with a
$1/\sqrt{\cal L}$ scaling.
This interesting behavior can be understood as follows: In principle,
the far detector measures some combination of the flux normalization
and $\stheta$. The turnover of the sensitivity line into the
second statistics dominated region occurs at the point, where the
determination of the normalization by the far detector itself becomes
more accurate than the external input $\sigma_\mathrm{norm}$. This
means that the limit becomes insensitive to the actual value of
$\sigma_\mathrm{norm}$. We illustrate this by the upper thin black
line, which shows the luminosity scaling for the case of larger
systematical errors. As an example, we choose values of
$\sigma_\mathrm{norm} = 1.7\%$ for the normalization and
$\sigma_\mathrm{cal} = 1.1\%$ for the energy calibration. We find that,
in this case, the transition to the systematics dominated regime occurs
at much smaller luminosities. However, for large luminosities, the same
 limit is approached as for the more optimistic case. The
diagonal gray curve shows the limit for no constraint at all on
the normalization and energy calibration.\footnote{Although we leave
the normalization free in the fit, we assume that the shape is known.}  Even in this extreme case, we obtain the same limit for high
luminosities.

This discussion demonstrates that the magnitude of the systematical error
determines the position of the sensitivity plateau, but
essentially does not affect the sensitivity at large
luminosities. We hence conclude that, for the case of the
\ReactorI\ setup, the systematical normalization error dominates. In order to
obtain a reliable limit, it should therefore be well under control. For large luminosities, such as for the \ReactorII\ setup, the sensitivity limit is independent of normalization errors because of the second statistics dominated regime. The robustness of this result will be further discussed in \Sec~\ref{sec:systematics}.

The horizontal line in \Fig~\ref{fig:luminosity} shows the typical
$\stheta$ sensitivity limit, which can be obtained from the first-generation \JHFSK\ superbeam experiment including correlations and degeneracies. It can be inferred that even the \ReactorI\ experiment gives comparable limits. The
competitiveness and complementarity of the information from reactor and superbeam
experiments will be discussed in greater detail in
\Secs~\ref{sec:RvsSB} and~\ref{sec:complementarity}. Finally, we note that with a
KamLAND-like detector of $1 \, \mathrm{kt}$ at a $10 \, \mathrm{GW}$ nuclear reactor with a running time of $\sim 5$ years, the high statistic region of a luminosity between $10^4 \, \mathrm{t \, GW \, y}$ and $10^5 \, \mathrm{t \, GW \, y}$ seems to be reachable.

\subsection{The impact of systematical errors}
\label{sec:systematics}

We discuss now the effects of various systematical errors
on the obtainable sensitivity to $\stheta$, where technical details are given
in \App~\ref{app:syst+ND}. Since it is a difficult task to estimate
systematical errors of a future experiment, our strategy has been to
find realistic values for various normalization errors (\eg, total
neutrino flux uncertainty, fiducial mass uncertainty) and energy
calibration errors. The chosen numbers are guided by the values obtained in existing reactor experiments, such as the
CHOOZ~\cite{Apollonio:1999ae,Apollonio:2002gd} or
KamLAND~\cite{Eguchi:2002dm} experiments. For other types of errors,
such as the shape uncertainty or the experimental systematical error,
we have adopted a conservative approach by choosing the worst case
situation of completely uncorrelated errors. In detail, we consider
the following effects (for the exact definitions, see
\App~\ref{app:syst+ND}):
\begin{enumerate}
\item\label{it:absnorm} We take into account a common overall
normalization error for the event rates of the
near and far detectors. Such an error could, for example, come from
the uncertainty of the neutrino flux normalization or the error on the
detection cross section. Typically, it is of the order of a few
percent.
\item\label{it:relnorm} We include uncorrelated normalization
uncertainties of the near and far detectors. Here contributes, for instance, the error on the fiducial mass of each detector. We assume that in this case an
error below 1\% can be reached.
\item\label{it:cal}
We take into account the energy calibration uncertainty by
introducing a parameter $g^A$ for each detector ($A=N,F$), and
replace the observed energy $E_\mathrm{obs}$ by
$(1+g^A)E_\mathrm{obs}$. We assume that the energy calibration is
known within an error of $\sigma_\mathrm{cal} \sim 0.5\%$.
\item\label{it:shape} In order to take into account an uncertainty of
the shape of the expected energy spectrum, we introduce an error
$\sigma_\mathrm{shape}$ on the theoretical prediction for each energy
bin which is completely uncorrelated between different energy bins.
This corresponds to the most pessimistic assumption of {\it no} knowledge
of possible shape distortions. However, we choose this error fully
correlated between the corresponding bins in the near and far
detector, since shape distortions should affect the signals in both
detectors of equal technology in the same way.
\item\label{it:exp} We include the possibility of an uncorrelated
experimental systematical error $\sigma_\mathrm{exp}$. Such an error
could, for example, result from insufficient knowledge of some
source of background. We call this uncertainty ``bin-to-bin error'' and take it
completely uncorrelated between energy bins, as well as between the
near and far detectors. Note that this corresponds again to the
worst case scenario, and values of $\sigma_\mathrm{exp}$ at the
per mill level should be realistic.
\end{enumerate}

As it is explicitely demonstrated in \App~\ref{app:syst+ND}, the overall
normalization error, the individual normalization errors of the two
detectors, and the energy calibration error of the near detector can be
merged into an effective normalization error $\sigma_\mathrm{norm}$
for the far detector.  Assuming realistic values of
2\%~\cite{Apollonio:2002gd} for the total normalization uncertainty and
0.6\%~\cite{oberauer} for the detector-specific uncertainty, we obtain
from \eqs~(\ref{equ:sigma_alpha}) and (\ref{equ:largeNN}) in
\App~\ref{app:syst+ND} an effective normalization error of
$\sigma_\mathrm{norm} = 0.8\%$, which is the standard
value for our numerical calculations.

\begin{figure}[ht!]
\begin{center}
\includegraphics[angle=-90, width=14cm]{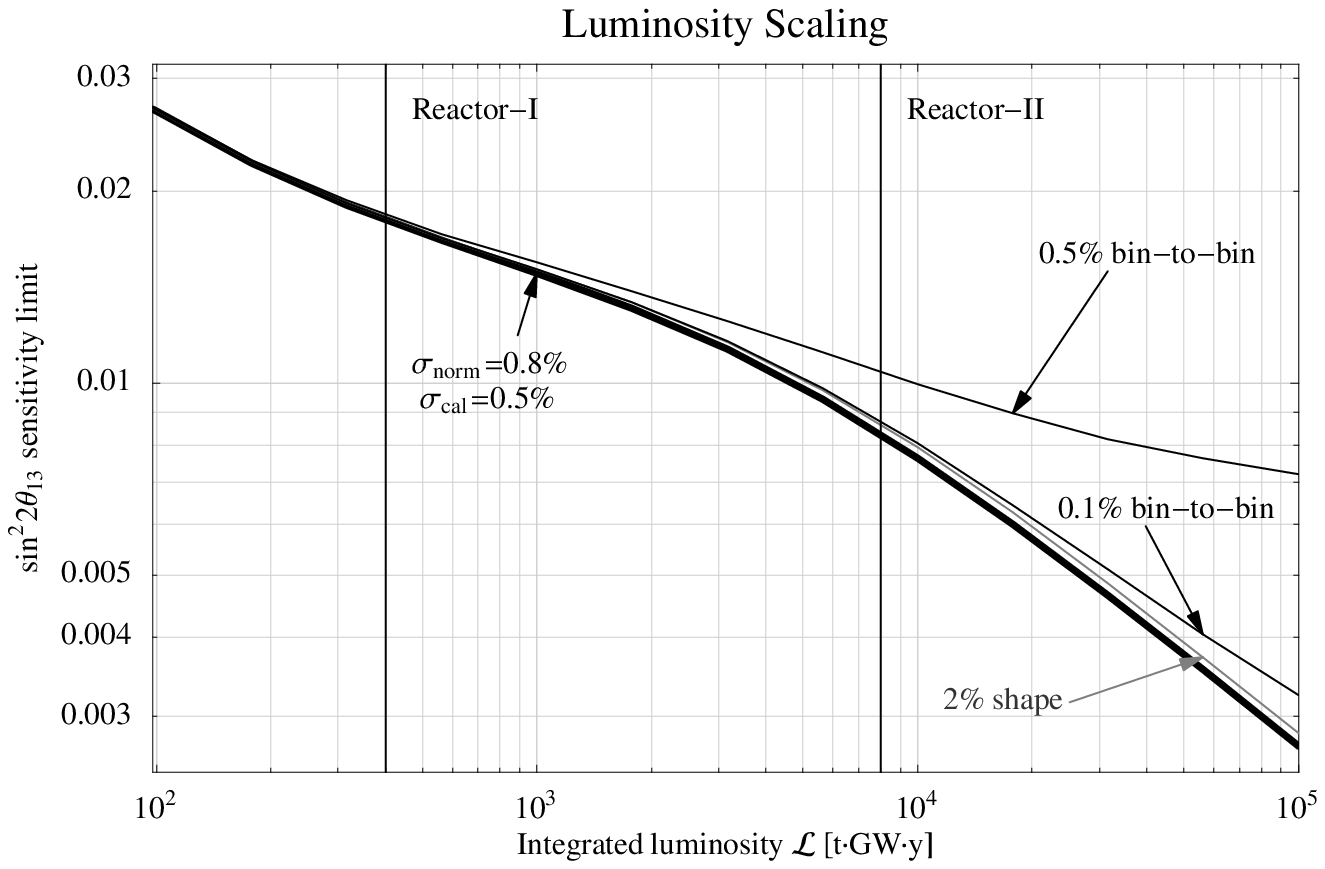}
\end{center}
\mycaption{\label{fig:systematics} The sensitivity to $\stheta$ as a function of the integrated luminosity for several selected error sources at the 90\% confidence level. The thick curve corresponds to our standard values for the errors on the flux normalization and the energy calibration. The gray curve has been  calculated by adding an error of 2\% for the shape of the predicted energy spectrum. For the curves labeled with ``bin-to-bin'', we include an uncorrelated experimental systematical error as given by the corresponding label. The vertical lines mark the integrated luminosities of our standard setups \ReactorI\ and \ReactorII\ as defined in \Tab~\ref{tab:reactor}.}
\end{figure}

The impact of a spectral shape uncertainty and of experimental
systematical errors is illustrated in \Fig~\ref{fig:systematics}, where
we show the luminosity scaling of the $\stheta$-limit for some values
of these errors.  The curve labeled with ``2\% shape'' corresponds to
an uncorrelated shape error on the predicted energy spectrum, such as
it is described in item~\ref{it:shape} above. This error covers a
large class of systematical effects, such as the uncertainty on the
$\beta$-decay spectrum of various isotopes in the reactor,
uncertainties of the fuel composition, or burn-up effects. As it can
be inferred from the figure, such an error has very little impact on
the sensitivity limit. The reason for this is that shape uncertainties
can be reduced very efficiently by the near detector, assuming that
there are at least 10 times more events in the near detector than in the far
detector. Our assumption of an uncorrelated error is the most
pessimistic theoretical error one could imagine. In more realistic
cases, one may expect some correlation of the shape uncertainty
between the energy bins, which could be even better eliminated by the
near detector.

The presence of a systematical, experimental bin-to-bin uncorrelated
error, as described in item~\ref{it:exp}, is potentially more
problematic for the measurement than any other error, since it cannot be
eliminated with the help of the near detector. As it is shown in
\Fig~\ref{fig:systematics}, such an error with a value below 0.5\% has
no effect for the \ReactorI\ setup, but it spoils the statistics
dominated region at high luminosities and would be important for large
detectors, such as \ReactorII . However, we note that 0.5\% for an
uncorrelated error of this type is a relatively large number, which should
be reducible down to the 0.1\% level. In this region,
the impact on the $\stheta$-limit becomes very small.  Suppose, for example,
that the number of background events in the detector is 1\% of
the reactor neutrino events. In this case, a 10\% knowledge of this
background is sufficient to reach a 0.1\% error on the
experimental event rate.

\subsection{Normalization and energy calibration errors, and the impact
  of an energy cut}
\label{sec:cut}

Since we have already demonstrated that the impact of an error on the expected
energy shape is very small, we will neglect it for the rest of this
work. Moreover, we further on assume that the experimental bin-to-bin
uncorrelated systematical error can be reduced to the
0.1\% level and hence can also be neglected. Then, as
shown in \App~\ref{app:syst+ND}, all systematical effects considered
in this paper can be reduced to two systematical errors, which are the
effective normalization uncertainty $\sigma_\mathrm{norm}$ and the
energy calibration error $\sigma_\mathrm{cal}$ in the far detector. We will discuss now in this subsection the impact of these two errors in
greater detail, combined with a possible low energy cut which could be
required in some cases to eliminate backgrounds.

\begin{figure}[ht!]
\begin{center}
\includegraphics[angle=-90, width=14cm]{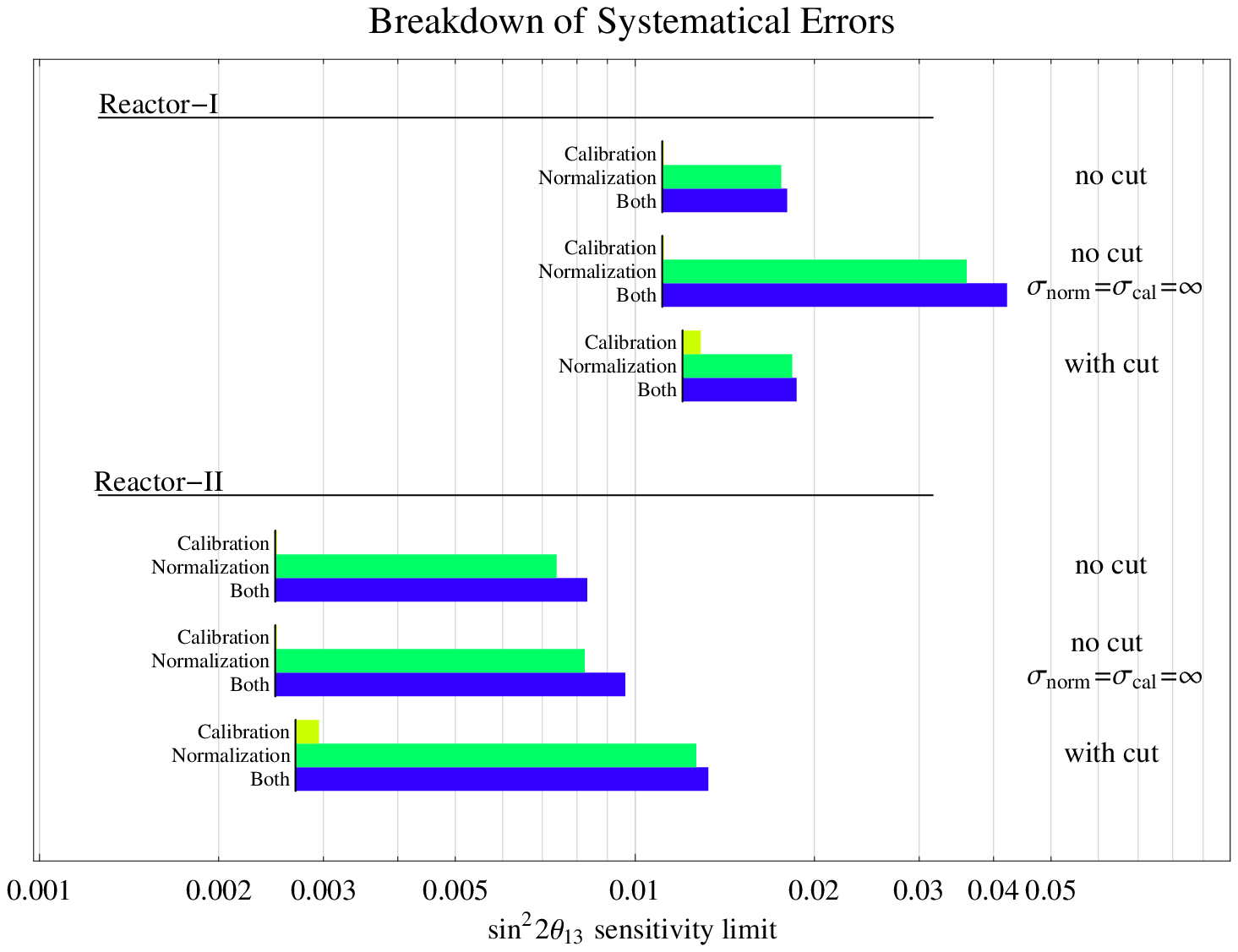}
\end{center}
\mycaption{\label{fig:mainimpacts} The sensitivity to $\stheta$ for the \ReactorI\ and \ReactorII\ setups at the 90\% confidence level. The left vertical line in each bar corresponds to statistical errors only, whereas the right edge shows the limit for different combinations of normalization
  and calibration errors. The sensitivity limits are shown for
  $\sigma_\mathrm{norm} = 0.8\%$ and $\sigma_\mathrm{cal}= 0.5\%$, or
  for $\sigma_\mathrm{norm} = \sigma_\mathrm{cal} \to \infty$ where
  indicated explicitly. Furthermore, the sensitivity limits are shown without cut on the energy, and for an energy cut at $E_\nu \ge 3.4 \, \mathrm{MeV}$.}
\end{figure}

In \fig~\ref{fig:mainimpacts}, we illustrate the $\stheta$ sensitivity
limits for various values of $\sigma_\mathrm{norm}$ and
$\sigma_\mathrm{cal}$. The vertical lines at the left edge of each bar correspond to the idealized case of statistical errors only. The bars indicate, how the limit deteriorates by including either the normalization error or the
calibration error or both. Let us first focus on the cases
labeled ``no cut'', where the full energy range of $1.8 \,
\mathrm{MeV} \le E_\nu \le \mathrm{8} \, \mathrm{MeV}$ is considered.
The final sensitivity limit is dominated by the normalization error,
and the calibration error has a rather small impact, although it is
not negligible. We also show in \figu{mainimpacts} the extreme case, where
$\sigma_\mathrm{norm}$ and $\sigma_\mathrm{cal}$ are set to infinity,
\ie, the event normalization and the energy scale are treated as free
parameters in the fit (\cf, \equ{chi2F} in \App~\ref{app:syst+ND}).  We learn
from this analysis that for $\sigma_\mathrm{cal}>0$ only, even if we
include the energy scale as a free parameter in the fit, the limit
is essentially the same as the pure statistics limit. This
demonstrates again that an energy scale uncertainty is not very
important. In addition, we observe that for \ReactorI\ the final
sensitivity limit is significantly worse for $\sigma_\mathrm{norm} =
\sigma_\mathrm{cal} \to \infty$ than for $\sigma_\mathrm{norm} =
0.8\%$, $\sigma_\mathrm{cal} = 0.5\%$, whereas the limits for
\ReactorII\ are essentially unchanged.  This again illustrates the
effect already discussed earlier in the context of
\fig~\ref{fig:luminosity}: At high luminosities the oscillation signal
is dominated by the spectral information and the far detector itself
provides an accurate measurement of the overall flux, which means that the
$\stheta$ sensitivity limit becomes insensitive to the value of
$\sigma_\mathrm{norm}$. On the contrary, for lower luminosities, such
as for the \ReactorI\ setup, a precise knowledge on
$\sigma_\mathrm{norm}$ is rather important, and the limit can be
significantly improved for small normalization errors.

\begin{figure}[ht!]
\begin{center}
\includegraphics[angle=-90, width=16cm]{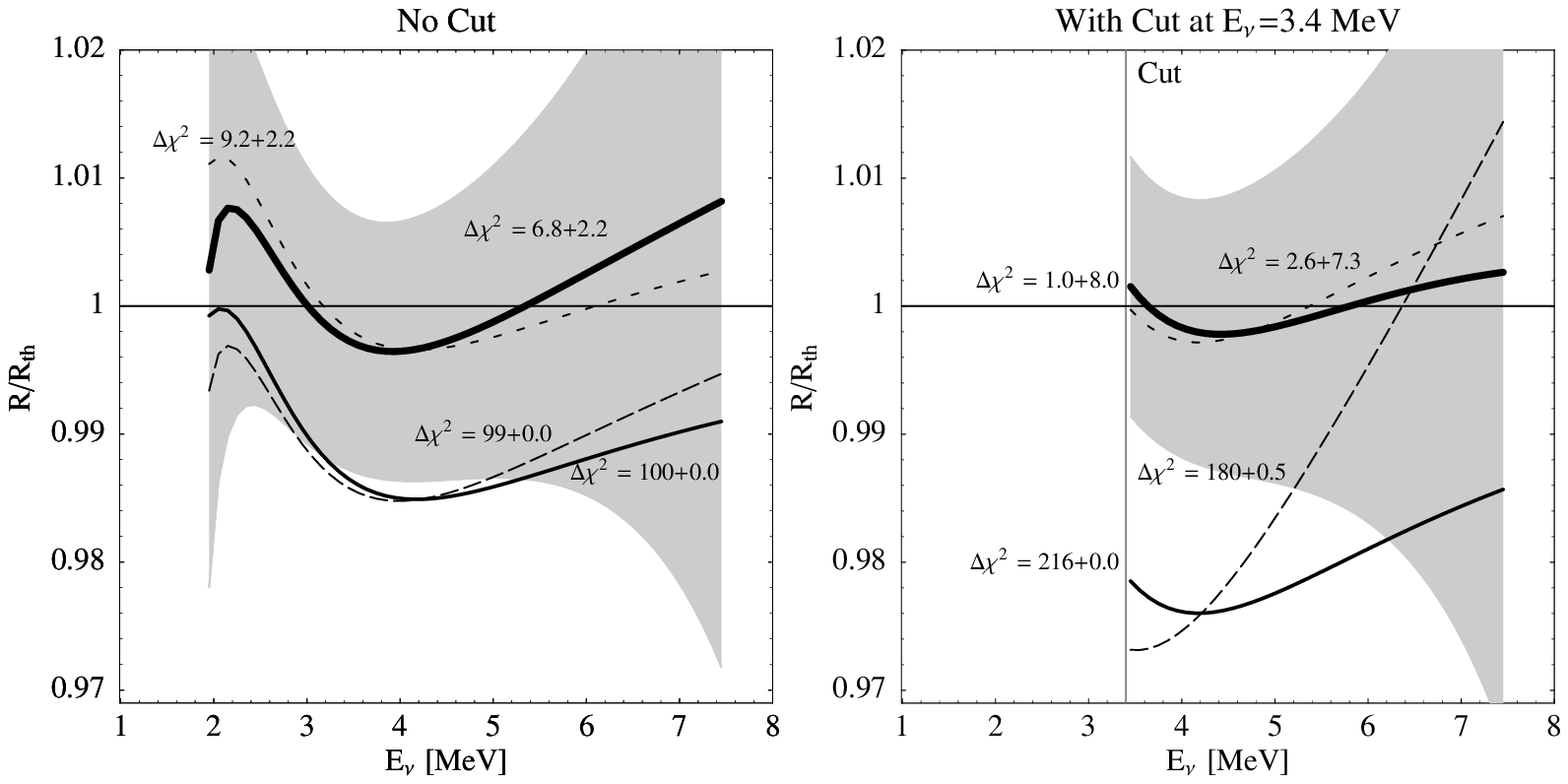}
\end{center}
\mycaption{\label{fig:shifted_rates} The ratio of the event rates at
  the $3 \sigma$ limit of $\stheta$ to the event rates for $\stheta =
  0$ for \ReactorII\ ($\stheta = 1.5 \cdot 10^{-2}$ in the left plot
  and $\stheta = 2.5 \cdot 10^{-2}$ in the right plot). In the
  left-hand panel, the full energy range is included, whereas in the
  right-hand panel, an energy cut at $E_\nu \ge 3.4 \, \mathrm{MeV}$
  is imposed. The thin solid curves correspond to the case of no
  systematical errors, whereas the long-dashed curves correspond to a
  calibration error of $\sigma_\mathrm{cal} = 0.5\%$, the short-dashed
  curves to a normalization error of $\sigma_\mathrm{norm} = 0.8\%$,
  and the thick solid curves to both errors switched on
  simultaneously. The shaded regions show the statistical errors, and
  the numbers given for each curve correspond to the separate
  contribution to the total value of $\Delta\chi^2$ of the data and
  the systematical errors, see \equ{chi2F}.}
\end{figure}

The cases in \figu{mainimpacts} labeled ``with cut'' correspond to the
situation when a low energy cut is imposed. Such a cut could be
motivated by some low energy background, such as from radioactivity of
the detector material~\cite{Apollonio:2002gd}.  For illustration, we
have chosen the cut $E_\nu \ge 3.4 \, \mathrm{MeV}$ corresponding to
$E_\mathrm{vis} \ge 2.6 \, \mathrm{MeV}$, as it is applied in the
KamLAND experiment to avoid the contribution of geo-neutrinos. Apart
from the slight worsening of the statistics-only limit due to smaller
event numbers, we find that the cut has a rather drastic impact for
the large detector \ReactorII.  The reason for this effect is
illustrated in \Fig~\ref{fig:shifted_rates}, where we show the
normalized energy spectrum for the following four cases: no
systematical errors, either $\sigma_\mathrm{norm}$ or
$\sigma_\mathrm{cal}$ included, and both $\sigma_\mathrm{norm}$ and
$\sigma_\mathrm{cal}$ simultaneously included.  One can see that the
cut at $3.4 \, \mathrm{MeV}$ is close to the oscillation
minimum at this baseline of $1.7 \, \mathrm{km}$. For such an
unfortunate choice, the interplay of normalization and calibration can
very efficiently reduce the oscillation pattern in the spectrum by
making the signal essentially flat. In fact, we have found that, in
the presence of an energy cut, our standard values for
$\sigma_\mathrm{norm}$ and $\sigma_\mathrm{cal}$ are sufficient to
completely destroy the statistics dominated regime at high
luminosities, similar to the ``bin-to-bin'' error shown in
\fig~\ref{fig:systematics}.
This problem can be avoided by making sure that the
oscillation minimum is safely included inside the accessible energy
range, such as for the case when no cut is applied. With this choice, it is impossible to destroy the oscillation signature by shifting the normalization or
stretching the energy scale, as it is clear from the left plot of
\Fig~\ref{fig:shifted_rates}. If a low energy cut cannot be avoided
because of backgrounds, one should eventually consider to change the
baseline in order to make sure that the oscillation minimum is covered by the
remaining energy window.

\section{The measurement of $\stheta$: reactors versus
  superbeams}
\label{sec:RvsSB}

In this section, we compare the potential to measure $\stheta$ in reactor experiments with the one of superbeams. As far as $\stheta$ is concerned, we will demonstrate that even the small reactor setup \ReactorI\ can compete with a first-generation superbeam, such as \JHFSK\ or \NUMI . There are essentially two interesting measurements for $\stheta$: the sensitivity to $\stheta$ and the precision of the measurement of $\stheta$. The discussion is therefore divided into two parts, where we also define the respective quantities.

\subsection{The sensitivity to $\stheta$}

We define the sensitivity or sensitivity limit to $\stheta$ as the largest value of $\stheta$, which fits the true value $\stheta \equiv 0$ at the chosen confidence level. For an experiment or a combination of experiments, it reflects the range between the sensitivity limit and the CHOOZ bound, where $\stheta>0$ could be detected.  With this definition, it should be clear that especially for superbeams the $\mathrm{sgn}(\Delta m_{31}^2)$-degeneracy has to be taken into account in the calculations, since the degenerate solution at $-|\Delta m_{31}^2|$ may allow a larger value of $\stheta$ fitting $\stheta \equiv 0$ than the best-fit solution. In addition, it has been demonstrated in \Ref~\cite{Huber:2002rs} that with this definition the sensitivity limits for the normal and inverted mass hierarchies are equal for the superbeams, since the zero rate vectors for the appearance channel are equal for $\stheta \equiv 0$. Similarly, \equ{PROBREACTOR} does not depend on the sign of $\Delta m_{31}^2$, which means that the sensitivity limit does not depend on the type of the hierarchy in this case, either. 

\begin{figure}[ht!]
\begin{center}
\includegraphics[width=16cm]{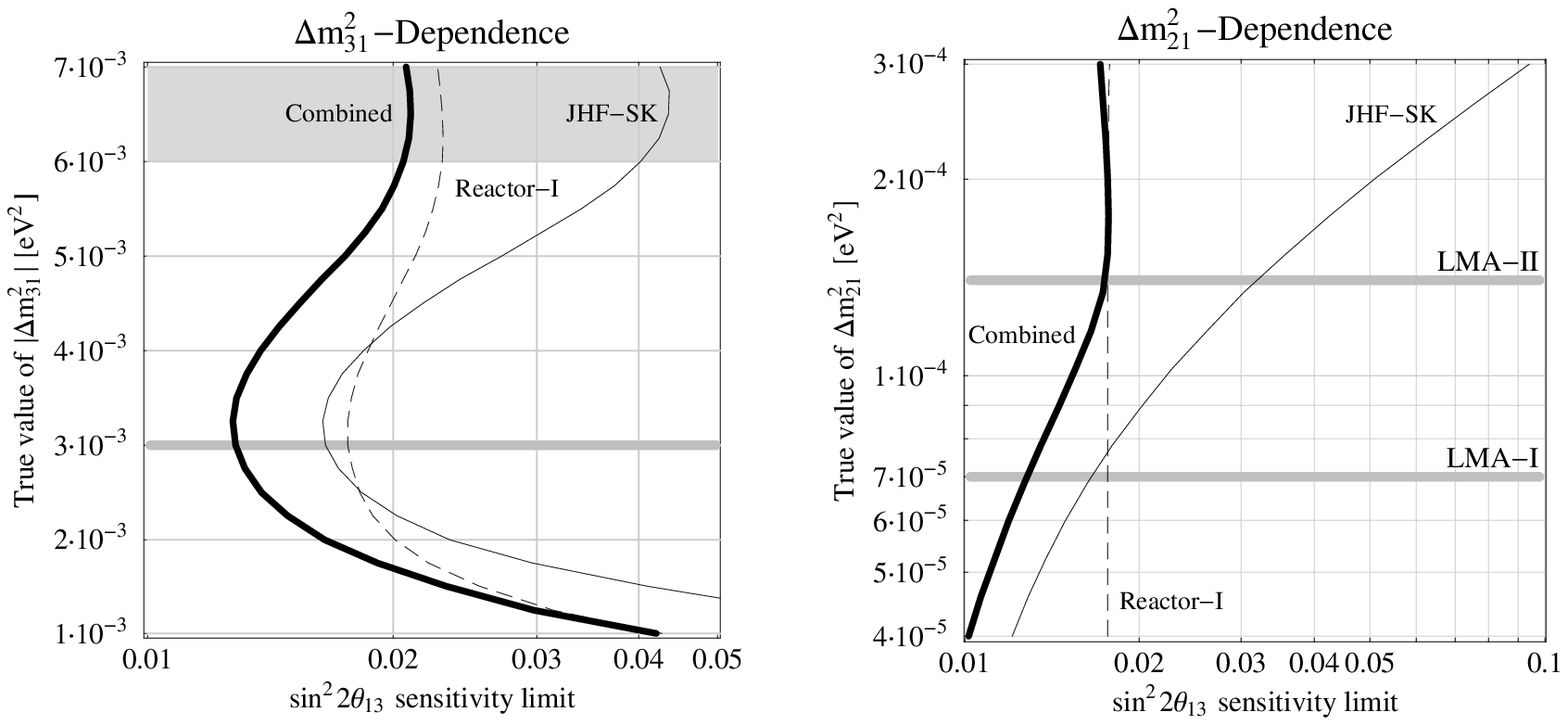}
\end{center}
\mycaption{\label{fig:dmdep} The sensitivity limits to $\stheta$ as
functions of the true values of $\Delta m_{31}^2$ (left plot) or
$\Delta m_{21}^2$ (right plot), respectively. They are shown for
the \JHFSK\ and \ReactorI\ setups as well as their combination at the
$90\%$ confidence level. In the left figure, the atmospheric
excluded region is shaded in gray, and in the right figure, only
the KamLAND-allowed region is shown. In both figures, the best-fit
values are marked by the gray lines.}
\end{figure}

\figu{dmdep} shows the sensitivity limits for \JHFSK\ as example for the superbeams and \ReactorI , as well as their combination at the $90 \%$ confidence level. In this figure, the left plot represents the $\ldm$-dependence and the right plot the $\sdm$-dependence. In order to compare experiments with similar capabilities, we have chosen the \ReactorI\ setups for these plots. The \ReactorII\ setup would be much better and would dominate the result. 

For the best-fit parameters used in this work, we obtain a sensitivity limit of $\stheta \simeq 1.8 \cdot 10^{-2}$ for \ReactorI\ at the $90 \%$ confidence level, as it can be read off from the figure. This result is in rather good agreement with the setups in \Refs~\cite{Minakata:2002jv} and~\cite{Martemyanov:2002td}, where similar values for the systematics uncertainties were used. In \figu{dmdep}, the $\ldm$-dependence clearly reflects the fact that both the \JHFSK\ and \ReactorI\ experiments are optimized for a value of $\ldm \simeq 3.0 \cdot 10^{-3} \, \mathrm{eV}^2$. However, as we have seen in \figu{basedep}, the broad reactor spectrum does not make this optimization peak as sharp as in the superbeam case, where the off-axis technology is actually used to produce a narrow-band beam. The risk of choosing a non-optimal value for the $\ldm$-optimization within the atmospheric allowed region is therefore much lower in the reactor case, though the superbeam is marginally better at the best-fit value. 
From the combination of the two experiments, we do not observe synergy effects which would improve the performance beyond a simple addition of statistics. As it is demonstrated in \fig~3 of \Ref~\cite{Huber:2002rs}, the \NUMI\ setup would lead to very similar results. 

As far as the $\sdm$-dependence is concerned, superbeams strongly suffer from the correlation with the CP phase. This can already be seen in \equ{PROBVACUUM}, where especially for large values of the hierarchy parameter $\alpha \equiv \sdm /\ldm$ the second and third terms become larger, and $\stheta$ becomes highly correlated with $\deltacp$. Thus, the \JHFSK\ setup loses almost any sensitivity to $\stheta$ below the CHOOZ bound for large values of $\sdm$ within the LMA-allowed region, as it can be seen in the right plot of \figu{dmdep}. On the other side, the reactor setup is hardly influenced by $\alpha$, since for the short baseline the effects of $\alpha$ within the LMA-allowed region are small. Thus, though a little bit worse for small values of $\sdm$, \ReactorI\ has a much better sensitivity than \JHFSK\ in most of the LMA-allowed region. The combination of the two experiments does, as in the case of $\ldm$, not show significant synergy effects beyond adding the statistics of the two experiments.

Apart from $\stheta$, the reactor experiment is sensitive to some parameters which can be more precisely measured by earlier experiments. We have therefore imposed external information on the parameters $\ldm$, $\sdm$, and $\sin^2 2 \theta_{12}$ appearing in \equ{PROBREACTOR}, and have studied the dependencies on the precision of this information. We find that the reactor experiment itself can measure the relevant parameters with a sufficient precision, as long as $\ldm$ is known better than to $30\%-50\%$ from external measurements. This required external precision of $\ldm$ is easily achievable by conventional beam experiments, such as  CNGS, MINOS, or K2K. The \ReactorI\ setup is in summary very competitive to the first-generation superbeams, especially since the risk of the unknown parameter values of $\ldm$ and $\sdm$ is considerably lower. With a higher luminosity, the \ReactorII\ setup could moreover do much better than the first-generation superbeams on comparable timescales.

\subsection{The precision of $\stheta$}

\begin{figure}[t!]
\begin{center}
\includegraphics[width=8cm]{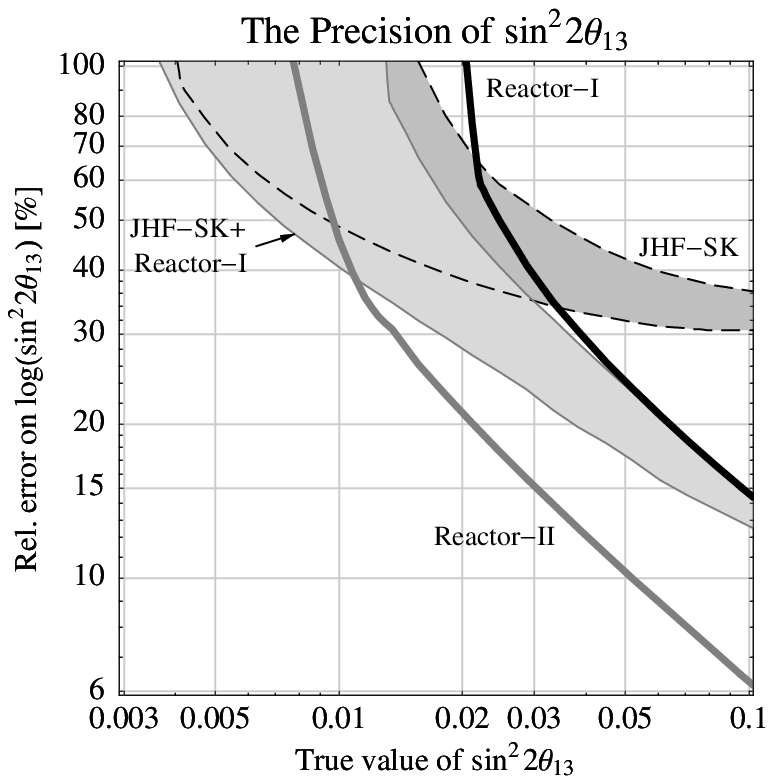}
\end{center}
\mycaption{\label{fig:th13prec} The relative error on $\mathrm{log}(\stheta)$ defined in \equ{defrelerror} as function of the true value of $\stheta$ for \JHFSK , \ReactorI , \ReactorII , and the combination of \JHFSK\ and \ReactorI\ at the $90\%$ confidence level. For experiments or combinations involving \JHFSK , the (by then unknown) true value of $\deltacp$ leads to bands representing all possible values of $\deltacp$. In addition, for each value of $\deltacp$, the worst among the best-fit and degenerate solutions was chosen.}
\end{figure}

As soon as $\stheta \neq 0$ is established by an experiment, the obtainable precision of $\stheta$ becomes interesting. This precision of the measurement of $\stheta$ depends, of course, on the true value of $\stheta$ itself. A meaningful quantity to discuss this precision is the relative error on $\mathrm{log} ( \stheta )$ as a function of $\stheta$~\cite{Huber:2002mx}, \ie,
\be
 \mathrm{ Rel. \, error \, on \,} \mathrm{log} (\stheta ) \, \, [ \% ] \equiv \frac{|\mathrm{log} (\stheta^{(2)}) - \mathrm{log} ( \stheta^{(1)}) |}{\mathrm{log} (\stheta )} \cdot 100.
\label{equ:defrelerror}
\ee 
Here $\mathrm{log} (\stheta^{(2)})$ and $\mathrm{log} ( \stheta^{(1)})$ refer to  the upper and lower intersections, respectively, of the $\chi^2$-function with the $90 \%$ confidence level. With this definition, it should be obvious that this relative error becomes larger than $100 \%$ close to the sensitivity limit. It is shown in \figu{th13prec} for \JHFSK\ and \ReactorI , their combination, and \ReactorII .
For any combination with a superbeam, we included in this calculation the degenerate solutions by taking the largest relative error of all degenerate solutions. In addition, we made no special assumptions about the unknown CP phase, which leads to the bands for experiments involving the superbeam in \figu{th13prec}. The lower edge of these bands corresponds to the best case, the upper edge to the worst case. This form of visualization takes into account that we do not know by then, which true value of $\deltacp$ has been realized by nature. The different results for different values of $\deltacp$ originate in the different shapes of the $\deltacp$-$\theta_{13}$-correlation, which means that this dependence does not necessarily imply that the experiment can measure $\deltacp$.

From \figu{th13prec}, we find that the precision of $\stheta$ is much
better for the reactor experiments for large values of $\stheta$, but
for small values of $\stheta$ the superbeams become better. This
behavior stems from the different nature of the signals: the reactor
experiments measure $\theta_{13}$ in the disappearance channel,
whereas the superbeams measure the appearance of events. For reactor experiments, the statistical and systematical errors are therefore
basically independent of $\stheta$, which implies that the precision of
$\stheta$ vanishes for small values of $\stheta$. On the
other hand, superbeams have, even for large values of $\stheta$, only a
limited number of signal events~\cite{Huber:2002mx}. Therefore, the
statistical errors are relatively large compared to a reactor
experiment. For smaller values of $\stheta$, however, the relative statistical error decreases as $1/\sqrt{N}$. This results in a
much better accuracy at low values of $\stheta$.  The same loss of precision for reactor experiments can also be
observed in the $\stheta$-$\ldm$-plane, as shown in
\fig~2 of \Ref~\cite{Minakata:2002jv}. This figure illustrates that
under a certain threshold value for $\stheta$ the shape of the
$\stheta$-$\ldm$ allowed region changes dramatically. Another
consequence of the shape of this region is that an external
measurement of $\ldm$ does not influence the precision of $\stheta$ in
\figu{th13prec} very much, because for large values of $\stheta$ the
parameters $\ldm$ and $\stheta$ are uncorrelated, and for small values
of $\stheta$ the relative error as defined in this section quickly
becomes very large before $\ldm$ and $\stheta$ become highly
correlated.

As far as the combination of \ReactorI\ and \JHFSK\ is concerned, there is some synergy for small values of $\stheta$. Though the precision of the reactor experiment deteriorates, both experiments together are somewhat better than the superbeam alone. The \ReactorII\ setup is however much better, even better than the combination of \ReactorI\ with \JHFSK\ for large values of $\stheta$, which means that a larger detector can really help in this case. After all, the superbeam suffers from the same problem as for the sensitivity limit: the larger $\sdm$ is, the worse the precision of $\stheta$ becomes. An unfortunate true value of the CP phase would additional erode the performance. Contrary to that, the \ReactorII\ setup does not have these problems, which means that a reactor experiment with a large detector would be the optimal choice to measure $\stheta$ with lower risks coming from oscillation parameter uncertainties.

\section{The complementarity of reactor experiments and superbeams}
\label{sec:complementarity}

We have already seen that the reactor experiments are very competitive in measuring $\stheta$. There are, however, other parameters, which the reactor experiments cannot access satisfactorily, such as the sign of $\Delta m_{31}^2$ and CP violation. We will therefore demonstrate now, how the reactor experiments would fit in the larger picture of the next-generation experiments.\footnote{A qualitative discussion of many interesting issues for the complementarity of future reactor and long-baseline experiments can also be found in \Ref~\cite{Minakata:2002jv}.} We will use the \ReactorII\ setup to identify the optimal combinations with the first-generation superbeams. This will demonstrate what reactor experiments can contribute in the best case.

\subsection{Precision measurements of the leading atmospheric parameters}

\begin{figure}[t!]
\begin{center}
\includegraphics[width=8cm]{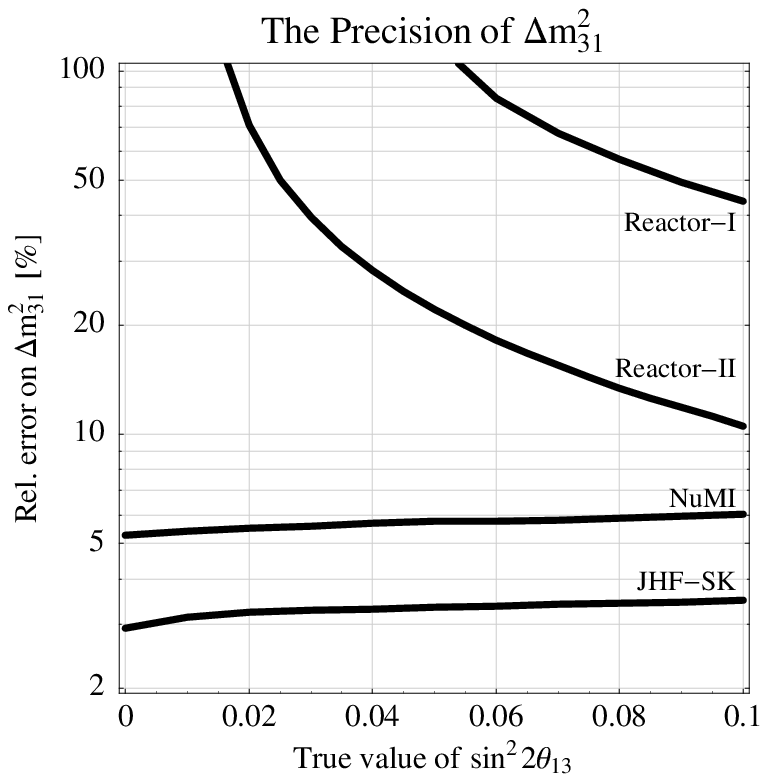}
\end{center}
\mycaption{\label{fig:dm31prec} The relative error on $\ldm$ in percent as function of the true value of $\stheta$ for \JHFSK , \NUMI , \ReactorI , and \ReactorII . For the calculation of the error, the full width of the $\chi^2$-function in the best-fit region is taken at the 90\% confidence level.}
\end{figure}

Among the interesting parameters for superbeams are the leading atmospheric parameters $\ldm$ and $\sin^2 2 \theta_{23}$. Compared to conventional beams, such as K2K, CNGS, and MINOS, the superbeams could achieve very high precisions for the leading atmospheric parameters. Reactor experiments are, at the short baselines we are considering, besides $\stheta$, only sensitive to $\ldm$. This can be understood in terms of \equ{PROBREACTOR}: the atmospheric mixing angle $\theta_{23}$ does not even appear in this equation and the sensitivity to the leading solar parameters can only be sensibly achieved at longer baselines, where $\Delta_{21}$ is large enough. Therefore, the first-generation superbeams would supply a measurement without serious competition at least for $\sin^2 2 \theta_{23}$. The atmospheric mass squared difference $\ldm$, however, could be measured by superbeams as well as reactor experiments, as it is illustrated in \figu{dm31prec}. There is nevertheless one important difference between those two types of experiments. The superbeams measure $\ldm$ with the disappearance channels, which are dominated by the leading atmospheric oscillation, and the measurement thus hardly depends on the true value of $\stheta$. For the reactor experiments, though, the parameter $\ldm$ is part of the signal proportional to $\stheta$. Therefore, they are strongly affected by the true value of $\stheta$, as it is demonstrated in \figu{dm31prec}. We hence conclude that long-baseline experiments are very important for precision measurements of the leading atmospheric parameters independent of the true value of $\stheta$.

\subsection{The sensitivity to the sign of $\ldm$}

\begin{figure}[t!]
\begin{center}
\includegraphics[width=8cm]{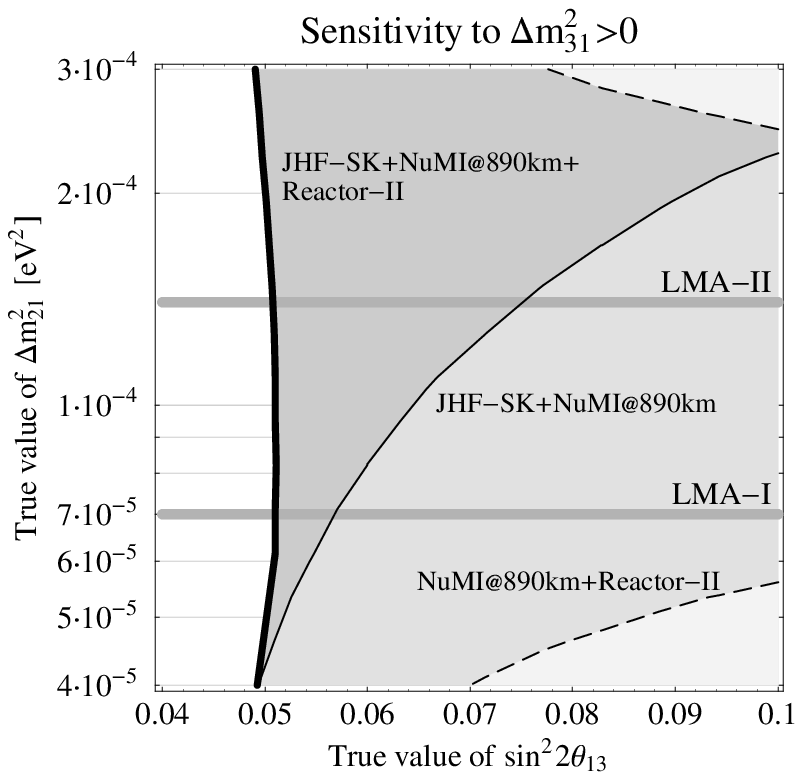}
\end{center}
\mycaption{\label{fig:sgndm31} The sensitivity to a positive sign of
$\ldm$ (normal mass hierarchy) as function of the true values of
$\stheta$ and $\sdm$ within the KamLAND-allowed region. Sensitivity at the 90\%
confidence level exists on the right-hand sides of
the curves. It is shown for several combinations of experiments: for
\JHFSK\ combined with \NUMI\ at a baseline of $890 \,
\mathrm{km}$~{\rm \protect\cite{Huber:2002rs}}, for \NUMI\ at a
baseline of $890 \, \mathrm{km}$ combined with \ReactorII , and for
\JHFSK\ combined with \NUMI\ at a baseline of $890 \, \mathrm{km}$ and
\ReactorII . The LMA best-fit values are marked by the horizontal gray
lines.}
\end{figure}

As it has been demonstrated in \Refs~\cite{Huber:2002mx,Huber:2002rs} and elsewhere, the mass hierarchy is, due to the $\mathrm{sgn}(\ldm )$-degeneracy, one of the hardest parameters to access for future long-baseline experiments. As it can be seen in \equ{PROBVACUUM}, the opposite sign of $\ldm$ opens, especially for large $\sdm$, the possibility of a degenerate solution at a different value of $\deltacp$, which makes it very hard to determine the mass hierarchy. We define that an experiment is sensitive to a certain sign of $\ldm$, if there is no possible solution with the opposite sign of $\ldm$ below the chosen confidence level. This sensitivity depends, similarly to the precision of $\stheta$, on the unknown true value of $\deltacp$. We take therefore the most conservative value of $\deltacp$ in order to show where sensitivity to the tested hierarchy exists independent of $\deltacp$. With this definition, it can be shown that neither the first-generation superbeams, such as \NUMI\ or \JHFSK , nor their combination, do have any sensitivity to the mass hierarchy~\cite{Huber:2002rs}. However, because of matter effects, which increase with the baseline, and the ability to resolve degeneracies with the combination of two superbeams~\cite{Barger:2002xk,Huber:2002rs}, the combination of \NUMI\ at a longer baseline together with \JHFSK\ is sensitive to the sign of $\ldm$ in a large region of the $\stheta$-$\sdm$-plane. The result for a positive sign of $\ldm$ and the combination of \JHFSK\ with \NUMI\ at a baseline of $890 \, \mathrm{km}$\footnote{This \NUMI\ baseline corresponds to the longest allowed baseline for the already fixed decay pipe at an off-axis angle of $0.72^\circ$.} from \Ref~\cite{Huber:2002rs} is, for example, shown in \figu{sgndm31}.
As it can be seen in the figure, the sensitivity to the normal mass hierarchy vanishes for large values of $\sdm$, which mainly comes from the $\mathrm{sgn}(\ldm )$-degeneracy.

We find however that the
{\it combination} of the \ReactorII\ setup at the short baseline of
$1.7 \, \mathrm{km}$ with {\em two} superbeam experiments significantly
improves the sensitivity to the mass hierarchy.  \figu{sgndm31} demonstrates that this combination (thick solid curve) has a very good sensitivity to $\ldm>0$ which does not depend on $\sdm$. The reactor experiment helps here indirectly to resolve the $\mathrm{sgn}(\ldm )$-degeneracy by the precision measurement of
$\stheta$. The combination of the three experiments thus leads to a
much better sensitivity to the normal mass hierarchy at large values
of $\sdm$ within the LMA-allowed region. In order to demonstrate that
one superbeam plus the reactor experiment at the short baseline is not enough to access the mass hierarchy, we show, in addition, in \figu{sgndm31} the
combination of the superbeam with the larger matter effects, \ie,
\NUMI\ at $890 \, \mathrm{km}$, together with \ReactorII . 
\figu{sgndm31} is for a normal mass hierarchy, but the inverted mass hierarchy produces rather similar results~\cite{Huber:2002rs}.

In \Ref~\cite{Petcov:2001sy}, it has been pointed out that for the case
of a relatively large $\sdm$, a reactor neutrino experiment at a
baseline of the order of 30 km might have some sensitivity to the sign
of $\ldm$ because of an interference term between $\sdm$ and $\ldm$
(see also \Ref~\cite{Schonert:2002ep}). This interference term is of third order proportional to $\alpha \cdot \stheta$, which means that it does not appear in our \equ{PROBREACTOR}. According to our numerical
results, a determination of the mass hierarchy at the $90\%$ confidence
level based on this effect is not possible, even for a combination of two
big reactor experiments with luminosities of $10^5 \, \mathrm{t \, GW \, y}$ at baselines of $1.7 \, \mathrm{km}$ and $30 \, \mathrm{km}$. 
In order to observe this tiny effect, the ratio $\sdm/\ldm$ and $\stheta$ should be as large as possible, and the solar mixing has to be far from
maximal mixing. Moreover, all these parameters have to been know with
(unrealistically) high precisions.

\subsection{The sensitivity to CP violation}

\begin{figure}[t!]
\begin{center}
\includegraphics[width=8cm]{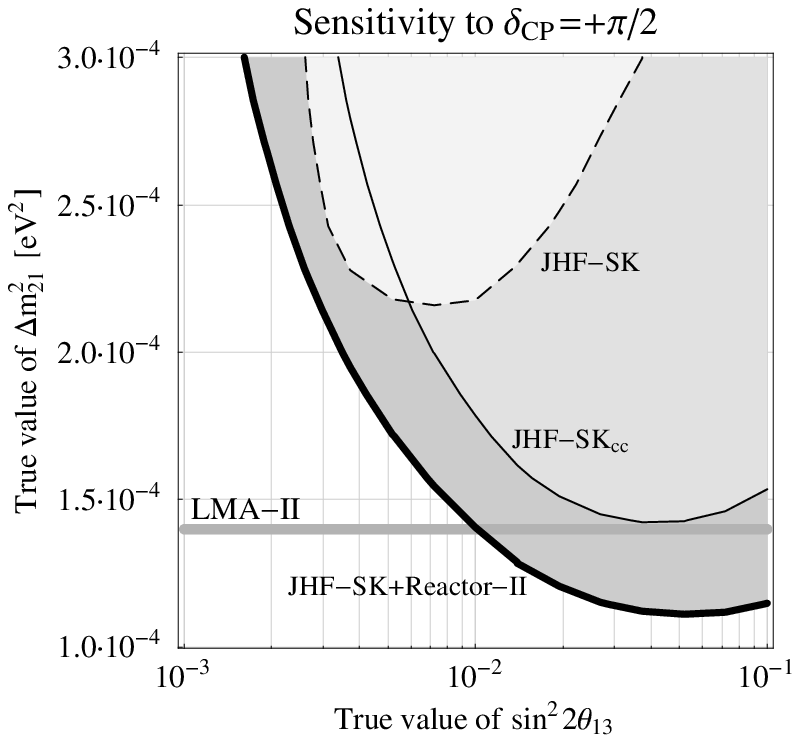}
\end{center}
\mycaption{\label{fig:cpviol} The sensitivity to maximal CP violation $\deltacp= + \pi/2$ as function of the true values of $\stheta$ and $\sdm$ within the KamLAND-allowed region. The sensitivity is at the 90\% confidence level given on the upper sides of the curves. It is shown for \JHFSK\ (neutrino running only), \JHFSK $_{cc}$ (same overall running time split into neutrino and antineutrino running with about equal numbers of events), and \JHFSK\ (neutrino running only) combined with \ReactorII . The LMA-II best-fit value is marked by the horizontal gray line.}
\end{figure}

There are various approaches to evaluate the capability of an experiment to access $\deltacp$. One of those, which is quite intuitive to understand and  qualitatively representative, is the sensitivity to maximal CP violation $\deltacp = \pm \pi/2$. In this spirit, we define that an experiment (or a combination) is sensitive to maximal CP violation, if the true value $\deltacp=\pm \pi/2$ does not fit the CP conserving values $0$ and $\pi$ at the chosen confidence level. With this definition, it is obvious that also the degenerate solutions have to be tested, since any degenerate solution with $\deltacp=0$ or $\pi$ fitting the maximal CP violation would destroy the sensitivity. It has been demonstrated in \Ref~\cite{Huber:2002rs} that for the first-generation superbeams the results qualitatively do not depend very much on the choice of the normal or inverted mass hierarchy and the true value $\deltacp=+\pi/2$ or $\deltacp=-\pi/2$. We show therefore only the results for the normal hierarchy and $\deltacp=+\pi/2$ in this work. In addition, it has been discussed in \Refs~\cite{Barger:2002xk,Huber:2002rs} and elsewhere that for the detection of CP violation with superbeams a combined neutrino and antineutrino running is very important -- either within one superbeam experiment or in the combination of different superbeams with different polarities. This result is not very surprising, since those two channels have a complementary dependence on the CP phase (\cf, \equ{PROBVACUUM}) and their combination helps to resolve the correlation between $\stheta$ and $\deltacp$. The antineutrino running has, however, one disadvantage: in order to accumulate similar event rates for comparable statistical weights, the antineutrino mode has to be operated much longer than the neutrino mode to compensate the lower cross section of antineutrinos. As illustrated for different setups in \figu{cpviol}, there is a possibility to circumvent this problem with \JHFSK . Instead of resolving the correlation between $\stheta$ and $\deltacp$ with the antineutrino channel, one could as well resolve it with a precision measurement of $\stheta$ with a reactor experiment.
The \JHFSK\ setup is in \figu{cpviol} shown in three different configurations: alone with neutrino running only (\JHFSK ), alone with the total running time split in order to obtain about equal numbers of neutrino and antineutrino events (\JHFSK $_{cc}$), and in combination with \ReactorII\ with neutrino running only (\JHFSK +\ReactorII ). The combined neutrino-antineutrino running at \JHFSK $_{cc}$ clearly helps to improve the performance compared to running with neutrinos only (\JHFSK ). There is, however, only a marginal improvement by adding the reactor experiment to \JHFSK $_{cc}$. The reason is that the correlation between $\stheta$ and $\deltacp$ is in this case already resolved by the antineutrino channel, and a precision measurement of $\stheta$ does not improve the sensitivity to CP violation further. The third, most interesting case in \figu{cpviol} shows the combination of \JHFSK\ with neutrino running only with \ReactorII . Because the statistics for \JHFSK\ is much better than for the combined neutrino and antineutrino running here, and the correlation between $\stheta$ and $\deltacp$ can be resolved by the reactor experiment, the overall performance is optimal. It even covers the LMA-II best-fit value, which leads to the interesting chance to observe leptonic CP violation by such a combination. We have not found any other combination (\eg, involving two superbeams as in \Ref~\cite{Huber:2002rs}), which could achieve such a good coverage in the $\stheta$-$\sdm$-plane by using the first-generation \JHFSK\ and \NUMI\ superbeams. Especially, the CP performance of \NUMI\ suffers from the fact that mainly the first term in \equ{PROBVACUUM} is enhanced by matter effects, which means that the relative weight of the second and third CP-sensitive terms is lower than in the \JHFSK\ case. A reactor experiment could thus be a very important element to measure $\deltacp$ without including an antineutrino running at superbeams. This is, however, because of the limited statistics,  for the first-generation superbeam experiments restricted to the LMA-II region.

\section{Summary and conclusions}
\label{sec:conclusion}

We have presented a detailed study of potential reactor
experiments with near and far detectors. The first important aspect
covered by this study has been a careful modeling and discussion of
the systematical errors of such reactor experiments. The second major
issue has been the comparison to the first-generation superbeams
\JHFSK\ (JHF to Super-Kamiokande) and \NUMI\, and the study of complementary effects between reactor experiments and superbeams. For a quantitative discussion, we have defined benchmark reactor experiments with
far detectors up to the size of the KamLAND detector, which we have
labeled as \ReactorI\ and \ReactorII . They have
integrated luminosities of $400 \, \mathrm{t \, GW \, y}$ and $8 \,
000 \, \mathrm{t \, GW \, y}$, respectively, where the units are given
as detector mass [tons] $\times$ thermal reactor power [GW] $\times$
running time [years].

\begin{figure}[ht!]
\begin{center}
\includegraphics[width=7.5cm,angle=-90]{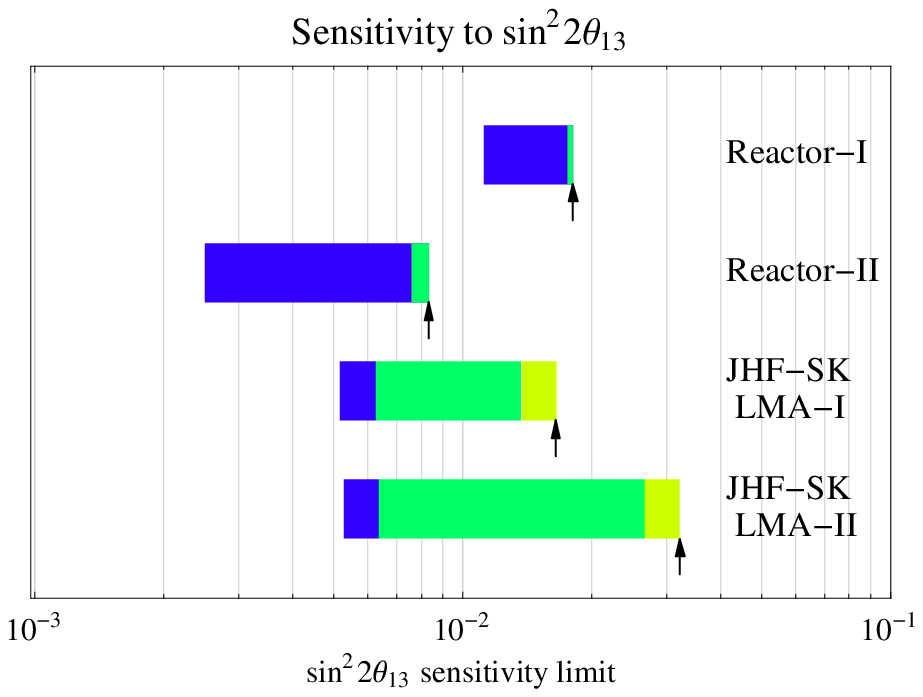}
\end{center}
\mycaption{\label{fig:summary} The sensitivity to $\stheta$ for \ReactorI , \ReactorII , \JHFSK\ at LMA-I ($\sdm =7 \cdot 10^{-5} \, \mathrm{eV}^2$), and \JHFSK\ at LMA-II ($\sdm =1.4 \cdot 10^{-4} \, \mathrm{eV}^2$) at the $90 \%$ confidence level. The sensitivitiy limits for the reactor experiments hardly depend on the true value of the solar parameters. The left edges of the bars correspond to the sensitivity limits from statistics only, the right edges of the bars to the real sensitivity limits after successively switching on systematics (dark/blue), correlations (medium gray/green), and degeneracies (light gray/yellow).}
\end{figure}

The structure of the relevant oscillation probabilities of reactor experiments demonstrates that they are only very little spoilt by correlations and not at all by degeneracies (\cf, \Sec~\ref{sec:framework}). Very good sensitivity limits for $\stheta$ can therefore be obtained. For our setups, we find, for the current best-fit values, sensitivity limits of $\stheta \simeq 1.8 \cdot 10^{-2}$ (\ReactorI ) and $\stheta \simeq 7.6 \cdot 10^{-3}$ (\ReactorII ), which is a significant improvement of the current CHOOZ bound. Compared to the superbeams, the \ReactorI\ sensitivity limit corresponds to the \JHFSK\ experiment including systematics, correlations, and degeneracies, which is about $\stheta \simeq 1.7 \cdot 10^{-2}$ for the LMA-I solution. The dependence of the sensitivity limits of \ReactorI , \ReactorII , and \JHFSK\ on systematics, correlations, and degeneracies is illustrated in \figu{summary}.
Compared to the superbeams, the sensitivities of the reactor experiments are almost unaffected by the true values of the solar parameters and less dependent on the value of $\ldm$. The dependence of the superbeam sensitivity limit on the LMA-I or LMA-II solution is shown in \figu{summary}, whereas the reactor experiments are hardly affected by $\sdm$. The sensitivity limits of the superbeams are, for very large $\sdm$ within the LMA-allowed region, not much better than the current CHOOZ bound. A reactor experiment is in contrast quite independent of the true values of the oscillation parameters. 

The careful modeling of the systematics leads to a number of important
 results.  First of all, an appropriate choice of the energy cuts
 turns out to be very important. Especially, the oscillation minimum
 should be within the energy window of the detectors, since spectral information is in reactor experiments very important for the reduction of systematical errors. The near and far detectors are assumed to be of identical technology in order to efficiently eliminate many systematical errors. It has turned out that the energy calibration error does not have a major impact on the results, since the spectral shape of the signal fixes the
 energy scale. The scaling of the $\stheta$ sensitivity limit with the integrated
 luminosity exhibits interesting features.  In the low luminosity
 regime, we find that the value of the normalization error is rather
 important. Thus, the \ReactorI\ setup would lose approximately $50\%$
 in sensitivity if the systematical errors are increased by a factor of
 $2$. At large integrated luminosities, such as for the case of
 our \ReactorII\ setup, this dependence on the systematical uncertainty
 of the normalization is however much weaker. The reason for this is that with a
 high statistics sample it is possible to simultaneously measure
 $\stheta$ \emph{and} the normalization with high precisions. Therefore, the
 $\stheta$ limit for the value of $\sigma_\mathrm{norm}=0.8\%$ is
 nearly the same as the one obtained with an unconstrained
 normalization, and, for luminosities above $8 \, 000 \, \mathrm{t \,
 GW \, y}$, the sensitivity limit again scales as
 $1/\sqrt{\mathcal{L}}$. At those large integrated luminosities,
 it becomes important if the calibration and normalization errors are sufficient to describe a real experiment, or if there are other sources of errors at the per mill level. We have therefore made a conservative estimate of the maximal
 admissible size of such additional error contributions and have found that an uncorrelated bin-to-bin error of $0.1\%$ would not spoil the
 $1/\sqrt{\mathcal{L}}$ behavior up to a integrated luminosity of at
 least $10^5 \, \mathrm{t \, GW \, y}$.  Increasing this uncorrelated
 bin-to-bin error by a factor of five to $0.5\%$, however, would lead
 to a saturation of the sensitivity limit at luminosities around $5 \,
 000 \, \mathrm{t \, GW \, y}$. For comparison, 
a detector of the KamLAND size at
a $10\,\mathrm{GW}$ power station with a running time of $5$ years would reach a luminosity of $\mathcal{L}=50 \, 000 \, \mathrm{t \, GW \, y}$. Those considerations indicate that a sensitivity to $\stheta$ of the order of $10^{-3}$ could be  feasible with an ambitious new reactor experiment.
  It however remains an open question at which level the systematics of a real experiment will finally put a hard limit on the sensitivity reach of a reactor 
experiment. A sensitivity increase of one order of magnitude beyond the current CHOOZ bound down to $\stheta=10^{-2}$ is however achievable quite independently of
the size of the systematical errors.

Even though reactor experiments are very good for $\stheta$ measurements, they cannot replace superbeams. Reactor experiments do not allow precision measurements of $\ldm$ independent of $\stheta$, and they are not at all sensitive to $\theta_{23}$. In addition, they do not have a significant  sensitivity to the mass hierarchy and cannot access $\deltacp$. Reactor experiments can, however, boost the performance of superbeams by the clean measurement of $\stheta$, which helps to resolve correlations and degeneracies. One could use, for example, a large reactor experiment together with \JHFSK\ in the neutrino running mode only, in order to obtain a better sensitivity to leptonic CP violation than with \JHFSK\ in a combined neutrino-antineutrino running mode. The reactor experiment resolves in this case the correlation between $\stheta$ and $\deltacp$, and the higher event rates of the neutrino running mode improve the overall performance. Another example is the determination of the mass hierarchy. Neither \JHFSK\ nor \NUMI\ alone could successfully determine the mass hierarchy. However, it has been shown in \Ref~\cite{Huber:2002rs} that their combination with a longer \NUMI\ baseline of $890 \, \mathrm{km}$ would be sensitive to the sign of $\ldm$ at least for small values of $\sdm$. It turns out that the \ReactorII\ setup together with the two superbeams in this configuration could, furthermore, be sensitive to the mass hierarchy quite independent of the true value of $\sdm$, because \ReactorII\ indirectly helps to resolve the $\mathrm{sgn}(\ldm )$-degeneracy by measuring $\stheta$ precisely. We do not observe this behavior with \ReactorII\ in combination with only one of the superbeams.

We conclude that the described reactor experiments are a very promising option on similar timescales to superbeams. Especially, the possibility to push the $\stheta$ sensitivity limit about one order of magnitude below the CHOOZ bound, down to $\stheta \simeq 10^{-2}$, is intriguing. This may be important, since most neutrino mass models, such as texture models, are linear in flavor space, and it would be very surprising if the diagonalization predicted extremely tiny values of $\stheta$. This means that the improvement of the CHOOZ bound achievable by reactor experiments would be very valuable to neutrino phenomenology and theory, and it would crucially influence the strategy of long-baseline discussion. For the superbeams, this implies that the task of finding $\stheta$ should be reconsidered, and that other observables, such as the mass hierarchy and CP violation, may get more weight in the planning. We have demonstrated that clever combinations of superbeams with a large reactor experiment may allow the determination of the mass hierarchy, and may even limit or measure $\deltacp$ with the first-generation superbeams. We therefore believe that the reactor discussion should be included in the optimization of superbeams.

%%%%%%%%%%%%%%%%%%%%%%%%%%%%%%%%%%%%%%%%%%%%%%%%%%%%%%%%%%%%%%%%%%%%
%%%%                      Acknowledgments                      %%%%
%%%%%%%%%%%%%%%%%%%%%%%%%%%%%%%%%%%%%%%%%%%%%%%%%%%%%%%%%%%%%%%%%%%%

\vspace*{7mm}
\subsubsection*{Acknowledgments}

We would like to thank L.~Oberauer and T.~Lasserre for very useful
discussions on experimental details of reactor experiments.  Further,
we thank S.T.~Petcov and S.~Choubey for comments to our work.

%%%%%%%%%%%%%%%%%%%%%%%%%%%%%%%%%%%%%%%%%%%%%%%%%%%%%%%%%%%%%%%%%%%%
%%%%                      Appendix                              %%%%
%%%%%%%%%%%%%%%%%%%%%%%%%%%%%%%%%%%%%%%%%%%%%%%%%%%%%%%%%%%%%%%%%%%%

\newpage

\begin{appendix}
\section{Systematical errors and the treatment of the near detector}
\label{app:syst+ND}

In this appendix, we describe in detail the $\chi^2$-analysis of the near-far
detector complex and how we implement the systematical errors discussed
in \Sec~\ref{sec:reactor_only}. 

We write for the theoretical prediction for the number of events in
the $i$th energy bin of the near ($A=N$) and far ($A=F$) detector,
respectively,
\begin{equation}\label{equ:theo}
  T^A_i = (1 + a + b^A + c_i) N_i^A  + g^A M_i^A \, ,
\end{equation}
and consider a $\chi^2$-function including the full spectral information
from both detectors:
\begin{equation}\label{equ:chi2N+F}
  \chi^2 = \sum_{A=N,F} 
  \left[ \sum_i 
      \frac{ (T_i^A - O_i^A)^2 }
           { O_i^A + \sigma_\mathrm{exp}^2 (O_i^A)^2 } +
    \left( \frac{b^A}{\sigma_b} \right)^2 +
    \left( \frac{g^A}{\sigma_\mathrm{cal}} \right)^2
  \right] +
    \sum_i  \left( \frac{c_i}{\sigma_\mathrm{shape}} \right )^2 +
  \left( \frac{a}{\sigma_a} \right)^2 \,.
\end{equation}
Here, $N_i^A$ is the expected number of events in the $i$th energy bin of the
corresponding detector, which depends on the oscillation parameters, and
$O_i^A$ is the observed number of events. We use $62$ bins in the range
between $E_{\bar\nu_e}=1.8\,\mathrm{MeV}$ and  
$E_{\bar\nu_e}=8.0\,\mathrm{MeV}$, corresponding to a bin width of 
$0.1\,\mathrm{MeV}$. In the absence of real data,
we take as ``observed number of events'' the expected number of events
for some fixed ``true values'' of the oscillation parameters. Per
definition, the near detector is as close to the reactor as that no
oscillations will occur, \ie, the $T_i^N$ do not depend on the
oscillation parameters, and we can set $O_i^N = N_i^N$. (We relax this
assumption in \App~\ref{app:ND}.)  

For each point in the space of oscillation parameters, the
$\chi^2$-function has to be minimized with respect to the parameters
$a,b^N$, $b^F$, $g^N$, $g^F$, and $c_i$ modeling the systematical
errors, such as described in \Sec~\ref{sec:systematics} in
items~\ref{it:absnorm} to~\ref{it:shape}. The parameter $a$ refers to
the error on the overall normalization of the number of events 
common to both detectors, and $\sigma_a$ is typically of
the order of a few percent. Furthermore, the parameters $b^N$ and
$b^F$ parameterize the uncorrelated normalization uncertainties of the
two detectors, where we assume that an error below 1\% can be reached.
The energy scale uncertainty in the two detectors is taken into account
by the parameters $g^N$ and $g^F$. To this aim we replace in $N_i^A$
the visible energy $E_\mathrm{vis}$ by $(1+g^A) E_\mathrm{vis}$. Then
we have to first order in $g^A$
\begin{equation}\label{equ:MA}
  N_i^A (g^A) \approx N_i^A(g^A = 0) + g^A \, M_i^A
  \quad\mbox{with}\quad 
  M_i^A = \left. \frac{\mathrm{d} N_i^A}{\mathrm{d} g^A} \right|_{g^A = 0} \,.
\end{equation}
A typical value for this error on the energy calibration is
$\sigma_\mathrm{cal} \sim 0.5\%$.  In order to model the uncorrelated
uncertainty on the shape of the expected energy spectrum, we introduce
in addition a parameter $c_i$ for each energy bin. Note that all the
parameters describing the systematical errors are at the percent
level, which means that the linear approximation in
\eqs~(\ref{equ:theo}) and (\ref{equ:MA}) is justified.  Finally, the
bin-to-bin uncorrelated experimental systematical error of
item~\ref{it:exp} in \Sec~\ref{sec:systematics} is included with the
help of the term containing $\sigma_\mathrm{exp}$ in \equ{chi2N+F}. In
this way we assume that the observed number of events in each bin and
each detector $O_i^A$ has in addition to the statistical error $\pm
\sqrt{O_i^A}$ the (uncorrelated) systematical error $\pm
\sigma_\mathrm{exp} O_i^A$.

In order to investigate the impact of the uncorrelated shape
uncertainty $\sigma_\mathrm{shape}$ and bin-to-bin experimental error
$\sigma_\mathrm{exp}$, we have performed an analysis using the complete
$\chi^2$ as given in \equ{chi2N+F}. However, as discussed in
\Sec~\ref{sec:systematics}, our numerical results are very insensitive
on the spectral shape uncertainty as long as $\sigma_\mathrm{shape}
\lesssim 2\%$. Therefore, it is save to set all of the $c_i$ to
$0$. Furthermore, we assume that $\sigma_\mathrm{exp} \lesssim 0.1\%$
can be reached, which means that setting this error to zero has little
impact on the numerical results. Being left with the 5 parameters
$a,b^N$, $b^F$, $g^N$, and $g^F$, we can analytically minimize
\equ{chi2N+F} (with $c_i=0$) with respect to the parameters $b^N$ and
$g^N$. The minimum is located at the values
\begin{equation}\label{equ:min}
b^N = a \, C_b  \,,\quad g^N = a \, C_g \, , 
\end{equation}
where
\begin{eqnarray}
C_b &=& -1 + \frac{1+\mathcal{M}'\sigma_\mathrm{cal}^2}
{(1+\mathcal{M}'\sigma_\mathrm{cal}^2)(1+\mathcal{N} \sigma_b^2) -
\mathcal{M}^2 \sigma_b^2 \sigma_\mathrm{cal}^2 } \, , \\
C_g &=& - \frac{\mathcal{M}\sigma_\mathrm{cal}^2}
{(1+\mathcal{M}'\sigma_\mathrm{cal}^2)(1+\mathcal{N} \sigma_b^2) - 
\mathcal{M}^2 \sigma_b^2 \sigma_\mathrm{cal}^2 } 
\end{eqnarray}
with the abbreviations
\begin{equation}
\mathcal{N} = \sum_i N_i^N \,,\quad
\mathcal{M} = \sum_i M_i^N \,,\quad
\mathcal{M}' = \sum_i \frac{ (M_i^N)^2 }{N_i^N} \,.
\end{equation}
Inserting \equ{min} into \equ{chi2N+F}, we
obtain an effective $\chi^2$-function for the far detector
\begin{equation}
  \chi^2_F = \sum_i 
    \frac{(T_i^F - O_i^F)^2}{O_i^F} +
    \left( \frac{b^F}{\sigma_b} \right)^2 +
    \left( \frac{g^F}{\sigma_\mathrm{cal}} \right)^2 +
    \left( \frac{a}{\bar\sigma} \right)^2 
\end{equation}
with
\begin{equation}\label{equ:sigmabar}
\frac{1}{\bar\sigma^2} =
\frac{1}{\sigma_a^2} + 
\left( \frac{C_b}{\sigma_b} \right)^2 +
\left( \frac{C_g}{\sigma_\mathrm{cal}} \right)^2 +
\sum_i \frac{ [ (1+C_b) N_i^N + C_g M_i^N ]^2 }{N_i^N} \,.
\end{equation}
Finally, we can take into account that only the sum $a+b^F$ appears in
the theoretical predictions in \equ{theo}. Therefore, introducing the new
parameter $\alpha = a+b^F$, we obtain
\begin{equation}\label{equ:chi2F}
  \chi^2_F = \sum_i 
    \frac{(T_i^F - O_i^F)^2}{O_i^F} +
    \left( \frac{g^F}{\sigma_\mathrm{cal}} \right)^2 +
    \left( \frac{\alpha}{\sigma_\mathrm{norm}} \right)^2 
\end{equation}
with
\begin{equation}\label{equ:sigma_alpha}
\sigma_\mathrm{norm}^2 = \bar\sigma^2 + \sigma_b^2\,.
\end{equation}
Hence, we have an effective $\chi^2$-function for the far detector which is given by \equ{chi2F}, where the information from the
near detector is properly taken into account by the error on the
normalization. The representative values $\sigma_\mathrm{norm} = 0.8\%$ and
$\sigma_\mathrm{cal} = 0.5\%$ used in our calculations should be
understood in the sense of \eqs~(\ref{equ:chi2F}), (\ref{equ:sigmabar})
and (\ref{equ:sigma_alpha}).

Let us eventually note that \eqs~(\ref{equ:sigmabar}) and
(\ref{equ:sigma_alpha}) show the correct behavior in all limiting
cases. For no near detector at all ($N_i^N,M_i^N \to 0$), we obtain
$C_b ,C_g \to 0$ and, as expected, $\bar\sigma \to \sigma_a$. For our
calculations, we assume the more interesting case of a very large
number of events in the near detector $\mathcal{N} \sigma_b^2 \gg
1$. In this limit, $C_b \approx -1, C_g \approx 0$ and
\begin{equation}\label{equ:largeNN}
\frac{1}{\bar\sigma^2} \approx 
\frac{1}{\sigma_a^2} + \frac{1}{\sigma_b^2} \,. 
\end{equation}
Assuming realistic values of $\sigma_a = 2\%$~\cite{Apollonio:2002gd}
for the flux uncertainty and $\sigma_b = 0.6\%$~\cite{oberauer} for
the detector-specific uncertainty, we obtain with
\eqs~(\ref{equ:sigma_alpha}) and (\ref{equ:largeNN}) an effective
normalization error of $\sigma_\mathrm{norm} \simeq 0.8\%$, which is
the value we have used for our numerical calculations.

%%%%%%%%%%%%%%%%%%%%%%%%%%%%%%%%%%%%%%%%%%%%%%%%%%%%%%%%%%%%%%%%%%%%%%%%%

\section{Experimental details of reactor experiments and summary
  of key assumptions}
\label{app:reactor}

The detector technology and performance of our proposed reactor
experiment is similar to the one of the CHOOZ~\cite{Apollonio:2002gd}
and KamLAND~\cite{Eguchi:2002dm} detectors, and the Borexino counting
test facility~\cite{Alimonti:1998nt}. Those detectors are based on a
sphere filled with a liquid scintillator, which is separated by a
plastic barrier from a buffer liquid between the sphere and the photo
multiplier tubes. The typical energy resolution for such a detector is
about $(5-10)\% /
\sqrt{E_\mathrm{vis}}$~\cite{Eguchi:2002dm,Schonert:2002ep}. We are
using an energy resolution of
$5\%/\sqrt{E_\mathrm{vis}}$~\cite{Alimonti:1998nt}, our results,
however, do not change for $7.5\%/\sqrt{E_\mathrm{vis}}$, which is
obtained for the KamLAND detector.  In addition, we assume that the
detector has a constant efficiency in the total analysis range from
the energy threshold at $E_\mathrm{vis}=1.0 \, \mathrm{MeV}$ up to
$E_\mathrm{vis}=7.2\,\mathrm{MeV}$, which we divide into $62$ bins
corresponding to a bin width of $0.1\,\mathrm{MeV}$. The normalization
is chosen such that the event rate per unit thermal power of the
reactor $\times$ fiducial detector mass $\times$ data taking time at a
distance of $1\,\mathrm{km}$ is given by $227.5 \, \mathrm{events}\,
\mathrm{t}^{-1}\, \mathrm{GW}^{-1}\,
\mathrm{y}^{-1}$~\cite{Schonert:2002ep}.

As shown in \Ref~\cite{Schonert:2002ep}, the accidental background
rate can be suppressed to one event per year for a $200\,\mathrm{t}$
detector, which is completely negligible for our purposes. Therefore,
we assume that it will be possible to construct a detector which is
basically free from backgrounds. In addition, cosmic backgrounds can
be efficiently rejected by using passive shielding of $\gtrsim
500\,\mathrm{mwe}$ ($\simeq 200\,\mathrm{m}$ rock overburden) together
with an active muon veto and pulse shape
discrimination~\cite{Schonert:2002ep}.  The number of events from
geo-neutrinos is at most $0.2\,\mathrm{events}\,\mathrm{t}^{-1}\,
\mathrm{y}^{-1}$~\cite{Schonert:2002ep}, which gives a total
contribution of at most $0.1\% $ to the total rate and should
therefore not present a limitation. For the short baselines used in
this work, the contribution of other power stations is very small.
Although some of the mentioned background sources may cause a total
contribution of the order of $1\%$, their actual influence will be
much smaller, since they are known to some extent and can also be
measured during the periods the reactor is switched off. Thus, a
$10\%$ determination of a $1\%$ background will contribute only to
$0.1 \%$ to the total error, and neglecting the backgrounds is an
excellent approximation for the far detector.

In our standard setups we assume a near detector which is close enough
to the reactor that no oscillations develop. (We relax this assumption in
\App~\ref{app:ND}.) In order to keep the characteristics of the near
and far detectors as similar as possible it may be necessary to use
exactly identical detectors. However, if it is possible to use a
smaller near detector it should have a size such that the event rate
is at least ten times higher than for the far detector without
oscillations. Of course, the near detector will have a much smaller
rock overburden than the far detector, and therefore the cosmic
background will be larger. However, because of the much larger event
rate in the near detector the signal to cosmic background ratio should
be even better than in the far detector, which means that we can also
neglect the backgrounds in the near detector. Let us illustrate this
by the following estimation: Consider, for example, our standard baseline of
$1.7\,\mathrm{km}$ for the far detector, a baseline of
$0.17\,\mathrm{km}$ for the near detector, and a near detector size
of a tenth of the far detector. For instance, with a rock overburden
of $50\,\mathrm{m}\simeq 125 \, \mathrm{mwe}$, the resulting muon flux
would be higher by a factor of $10$~\cite{Apollonio:2002gd}.  Since
the near detector has about a tenth of the volume of the far detector,
its surface area is about $10^{2/3}\simeq 4.6$ smaller. Therefore, the
total number of muon induced events is approximately twice as high as
the one of the far detector. However, since the near detector has
about ten times as many signal events as the far detector, the signal
to cosmic background ratio should be a factor of $5$ better than in
the far detector, and neglecting the background is an excellent approximation.

In the following we summarize the most important assumptions about the
reactor neutrino experiment adopted in our calculations:
\begin{itemize}
\item
We consider one single reactor block.
\item
We assume that neglecting backgrounds is a good approximation for the
near as well as for the far detector (see \Ref~\cite{Schonert:2002ep}
and the estimates above).
\item
The full energy spectrum above the threshold is used. The impact
of a low energy cut is investigated in \Sec~\ref{sec:cut}.
\item
Detection efficiencies are assumed to be constant in the full energy
interval. 
\item
To convert the integrated luminosity $\mathcal{L} =$ 
fiducial detector mass [tons] $\times$ thermal reactor power [GW] $\times$
running time [years] into the number of events we follow the rule
given in \Ref~\cite{Schonert:2002ep}, assuming
\begin{itemize}
  \item a PXE-based scintillator, 
  \item the reactor running full time at nominal thermal power,
  \item 100\% detection efficiency.
\end{itemize}
Other detector materials, a lower efficiency, reactor-off periods, or
an operation at a lower thermal power, lead to a simple rescaling of our
results.
\item
We assume that the two detectors can be kept relatively calibrated
during the full measurement period.
\item
In our standard setups the near detector is situated as close
($\lesssim 200 \, \mathrm{m}$) to the reactor as no oscillations
develop. The impact of larger near detector baselines is investigated
in \App~\ref{app:ND}.
\item
Finally, our assumptions about systematical errors and their relevance
are summarized in \Tab~\ref{tab:error_summary}.
\end{itemize}  

\begin{table}[h!]
\begin{center}
\begin{tabular}{|ll|c|c|}
\hline
Standard assumption & &\ReactorI\ &\ReactorII\ \\
\hline
Effective normalization  & $\sigma_\mathrm{norm} = 0.8 \%$ &
important & not important\\
Energy calibration & $\sigma_\mathrm{cal} = 0.5\%$ &
not important & not important\\
Exp.~bin-to-bin uncorr.~error & $\sigma_\mathrm{exp} \lesssim 0.1\%$ &
not important & important\\
\hline
\end{tabular}
\end{center}
\mycaption{\label{tab:error_summary} Standard assumptions about
  systematical errors and their relevance for the two reactor
  benchmark setups \mbox{\ReactorI} and \ReactorII\ used in this
  work.}
\end{table}

\section{The position of the near detector}
\label{app:ND}

For practical reasons it might be hard to find a reactor station
where a near detector can be situated very close ($\lesssim 200 \,
\mathrm{m}$) to the core with sufficient rock overburden. Therefore,
it is interesting to investigate the impact of larger near detector
baselines on the $\stheta$ limit. In this case there will be already
some effect of oscillations in the near detector and it is not possible 
to simplify the analysis as discussed in \App~\ref{app:syst+ND}. 
Therefore, we have performed an analysis based on the full
$\chi^2$-function given in \equ{chi2N+F}, taking into account the
effect of oscillations in both detectors. Now the information
provided by the near detector on the initial flux normalization and
energy shape is already mixed with some oscillation signature. Hence,
one expects that the correct treatment of the shape uncertainty
$\sigma_\mathrm{shape}$ due to the coefficients $c_i$ in \equ{chi2N+F}
becomes more important.

\begin{figure}[ht!]
\begin{center}
\includegraphics[width=14cm]{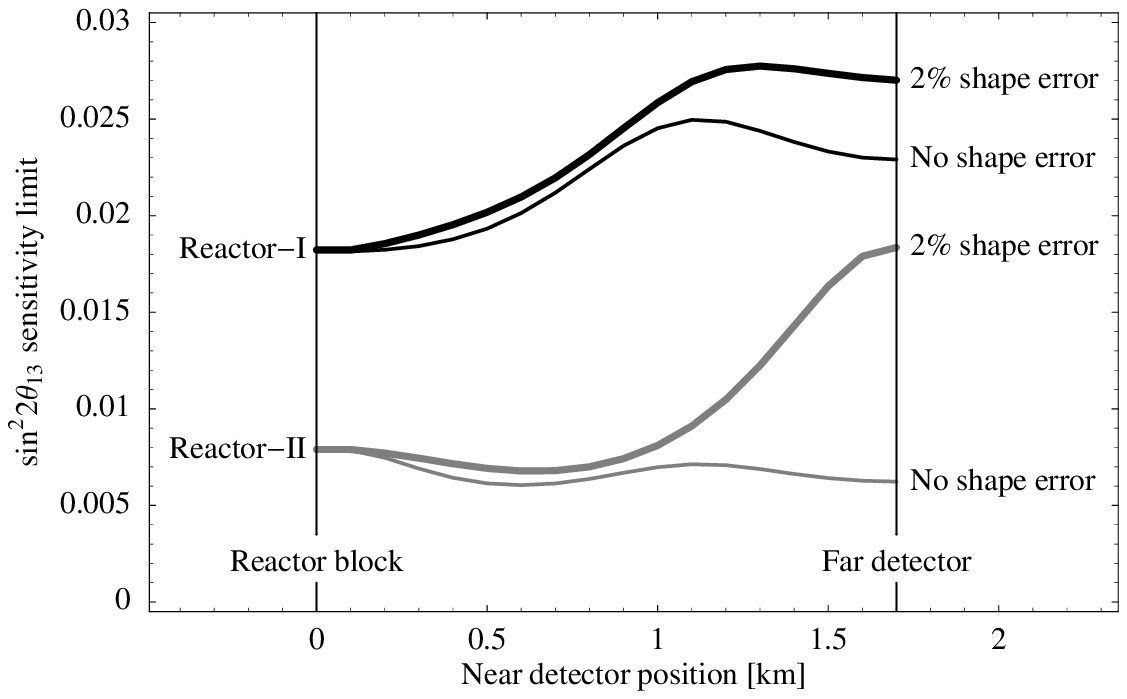}
\end{center}
\mycaption{\label{fig:neardet} The sensitivity to $\stheta$ at the $90
  \%$ confidence level for \ReactorI\ and \ReactorII\ as a function of
  the near detector position. The far detector is situated at 1.7~km
  and we assume identical detectors. Furthermore, the impact of an
  uncorrelated theoretical shape uncertainty of 2\% is shown.}
\end{figure}

The results of this analysis are presented in
\Fig~\ref{fig:neardet}. We find that for the case of \ReactorI\ the
limit starts deteriorating around a near detector distance of 400 m,
whereas for \ReactorII\ the limit even improves slightly up to near
detector baselines of $\sim 1 \, \mathrm{km}$. Due to the high
statistics in the case of \ReactorII, flux normalization and shape are
very well determined by the near detector even in the presence of some
effect of $\stheta$, and the additional information on oscillations
improves the limit a bit. Furthermore, we find from
\Fig~\ref{fig:neardet} that the shape uncertainty becomes important
for near detector baselines $\gtrsim 1.1 \, \mathrm{km}$, especially
for \ReactorII. A reduction of this theoretical error would be helpful
in such a situation.  We note that assuming the shape error to be
completely uncorrelated corresponds to the worst case. A more
realistic implementation of the shape uncertainty including correct
correlations will lead to results somewhere in between the curves for
no and 2\% shape error in \Fig~\ref{fig:neardet}. These calculations
have been done for our standard value of $\ldm = 3 \cdot 10^{-3}
\,\mathrm{eV}^2$. However, we find very similar behavior for other
values of $\ldm$.

To summarize, for the case of \ReactorI-like experiments one should
look for a site where the near detector can be placed at a distance of
at most 400 m from the reactor. For large detectors, such as
\ReactorII, near detector baselines of up to 1 km will perform
well. For near detector baselines longer than about 1 km the correct
treatment of the theoretical shape uncertainty becomes important.

\end{appendix}

%%%%%%%%%%%%%%%%%%%%%%%%%%%%%%%%%%%%%%%%%%%%%%%%%%%%%%%%%%%%%%%%%%%%%
%%%%                       References                            %%%%
%%%%%%%%%%%%%%%%%%%%%%%%%%%%%%%%%%%%%%%%%%%%%%%%%%%%%%%%%%%%%%%%%%%%%

\end{document}